\documentclass[pra,aps,superbib,citeautoscript,twocolumn]{revtex4-1}

\usepackage[colorlinks=true,urlcolor=blue,citecolor=blue,linkcolor=blue]{hyperref}

\usepackage{amsmath,bm}
\usepackage{amstext}
\usepackage{epsfig}
\usepackage{xcolor}
\usepackage{subfig}
\usepackage{graphicx,amsmath,amssymb,tabularx}
\usepackage{multirow}
\usepackage{array}
\usepackage{dsfont}
\usepackage{caption}
\usepackage{cleveref}
\usepackage{ulem}
\usepackage[toc]{appendix}
\usepackage{physics}
\usepackage{mathrsfs}  
\usepackage{amsbsy}

\definecolor{dgreen}{rgb}{0,.5,0}
\definecolor{dred}{rgb}{.7,.0,.0}

\DeclareMathAlphabet\mathbfcal{OMS}{cmsy}{b}{n}

\newcommand{\br}{\mathbf{r}}

\newcommand{\ie}{{\it i.e.}}

\newcommand{\be}{\begin{eqnarray}}
\newcommand{\ee}{\end{eqnarray}}
\DeclareMathAlphabet\mathbfcal{OMS}{cmsy}{b}{n}

\newcommand{\rev}[1]{{{#1}}}
\newcommand{\revben}[1]{{{#1}}}

\begin{document}

\title{
Density functional theory beyond the Born--Oppenheimer
approximation: \rev{Exact mapping onto an electronically non-interacting Kohn--Sham molecule}}

\author{
Emmanuel Fromager$^1$ and Benjamin Lasorne$^2$
}
\affiliation{\it 
~\\
$^1$Laboratoire de Chimie Quantique,
Institut de Chimie, CNRS / Universit\'{e} de Strasbourg,
4 rue Blaise Pascal, 67000 Strasbourg, France\\
\\
$^2$
ICGM, Univ Montpellier, CNRS, ENSCM, Montpellier, France
}


\begin{abstract}
\rev{This work presents an alternative, general, and in-principle exact extension of electronic Kohn--Sham density functional theory (KS-DFT) to the fully quantum-mechanical molecular problem. Unlike in existing multi-component or exact-factorization-based DFTs of electrons and nuclei, both nuclear and electronic densities are mapped onto a fictitious electronically non-interacting molecule (referred to as KS molecule), where the electrons still interact with the nuclei. Moreover, in the present molecular KS-DFT, no assumption is made about the mathematical form (exactly factorized or not) of the molecular wavefunction. By expanding the KS molecular wavefunction {\it à la} Born--Huang, we obtain a self-consistent set of ``KS beyond Born-Oppenheimer" electronic equations coupled to nuclear equations that describe nuclei interacting among themselves and with non-interacting electrons.} An exact adiabatic connection formula is derived for the
Hartree-exchange-correlation energy of the electrons within the molecule and, on that basis, a practical adiabatic density-functional approximation is
proposed and discussed. 
\end{abstract}

\maketitle



\section{Introduction}

\rev{Electronic} Kohn--Sham density functional theory (KS-DFT)~\cite{Hohenberg1964,Kohn1965} and its extension to the time-dependent linear response regime~\cite{runge1984density,Casida_tddft_review_2012} have become over the last couple of decades the workhorse of quantum chemistry and materials science~\cite{Teale2022}. Applied within the Born--Oppenheimer (BO) approximation, they enable to calculate, in principle exactly, both electronic ground-state and vertical excitation  energies, respectively, in a given molecular geometry. \rev{In this work, we focus on situations where the BO approximation breaks down}, for example, in the vicinity of a conical intersection \rev{\cite{bae06,dom04,dom11,las11:460}}, which is a low-symmetry generalization of the electronic degeneracy responsible of the infamous Jahn-Teller effect. In this case, both quantum nuclear effects and nonadiabatic nuclear-electron correlations (so-called vibronic nonadiabatic couplings) may be significant -- almost always when addressing a manifold of excited electronic states, but also sometimes when targetting the molecular (vibronic)  ground state -- and, therefore, they should be incorporated into the \rev{density-functional description of the electronic structure within the molecule}.\\ 

\rev{This issue has been addressed in the past in different ways}. Let us first mention Multicomponent DFT (MCDFT)~\cite{Kreibich2001,Gidopoulos98,Butriy07,Kreibich08} (see also Refs.~\onlinecite{Chakraborty2008_Development,delaLande2019_Multicomponent,Xu2023_First-Principles} and the references therein). In MCDFT, the KS system \rev{is a molecular system that} consists of non-interacting electrons (\ie, electrons that do {\it not} interact among themselves) and of nuclei that interact {\it only} among themselves~\cite{Kreibich08}. \rev{More recently, a quite different DFT of electrons and nuclei~\cite{Requist16_Exact,Li18_Density}, which is based on the well-established exact factorization (EF) of the molecular wavefunction~\cite{Hunter1975_Conditional,Abedi10_EF,Abedi12_Correlated,Gidopoulos2014_Electronic,Min15_Coupled} (we refer the reader to Ref.~\cite{Villaseco22_Exact}, which gives an updated and comprehensive overview of the EF approach in various realms, and the references therein), has been proposed and explored.} \rev{This extension of the EF formalism from many-electron wavefunctions to a KS formulation is referred to as EF-DFT in the following}. While mapping the electronic conditional density (and the paramagnetic current density) onto a non-interacting KS electronic system, the true physical marginal nuclear wavefunction is still evaluated, in principle exactly, in such a scheme~\cite{Requist16_Exact}. A very appealing feature of the EF formalism is the possibility that it offers to develop quantum-classical approaches~\cite{Min15_Coupled-Trajectory,Talotta2020_Spin-Orbit,Villaseco22_Exact} from first principles. Its combination with ensemble DFT also gave promising results~\cite{Filatov18_Direct}.\\

\rev{In this work, we follow a different path, by adopting, as a starting point, an electronic KS-DFT viewpoint on the fully quantum-mechanical molecular problem. In standard beyond-BO simulations, KS-DFT and linear response TDDFT calculations are performed for fixed molecular geometries, thus giving access to electronic (ground- and excited-state) energies and non-adiabatic couplings~\cite{Wang21_NAC-TDDFT,Send10_First-order,Ou15_First-order}, from which the coupled nuclear Schr\"{o}dinger equations can be solved. Bypassing the (computationally more involved than a KS-DFT calculation) linear response TDDFT step, while preserving the simplicity of the standard (electronically non-interacting) KS scheme, is highly desirable for practical applications. This is the original motivation for formulating an alternative, general, and in-principle exact extension of electronic KS-DFT to the beyond-BO description of molecules. For that purpose, we proceed by analogy with regular electronic KS-DFT and scrutinize the nuclear-electron attraction potential energy (evaluated for quantum electrons and nuclei) from which two basic variables naturally emerge. The first one is an effective electronic density which is defined for any mathematical form of the molecular wavefunction (exactly-factorized or not), unlike in EF-DFT, and which is geometry-dependent, unlike in MCDFT. The second variable is the nuclear density. On that basis, we derive what we refer to as molecular KS-DFT, where the full molecular problem is mapped onto a fictitious electronically non-interacting molecule (referred to as KS molecule). When the molecular wavefunction of the latter (so-called KS molecular wavefunction) is Born--Huang (BH) expanded~\cite{bor54}, we obtain the desired ``KS beyond BO" scheme, where KS-like equations, which incorporate (through both nuclear and electronic densities) the quantum description of the nuclei beyond the BO approximation, are coupled to nuclear equations that describe nuclei interacting among themselves and with non-interacting electrons. The self-consistent and coupled solutions to the electronic and nuclear equations reproduce, in principle exactly, both electronic and nuclear physical densities. The exact  ground-state molecular energy can then be determined through the analog for the full molecular problem of the electronic Hartree-exchange-correlation (Hxc) density functional.\\
}

\rev{The paper is organized as follows. After a brief review of electronic KS-DFT in Sec.~\ref{sec:review_elec_KSDFT}, we motivate our choice of basic variables in Sec.~\ref{sec:identification_basic_var}. On that basis, we formulate the molecular KS-DFT in Sec.~\ref{sec:KS_mol_DFT}, and discuss its practical implementation based on the BH expansion of the KS molecular wavefunction in Sec.~\ref{sec:mol-KS_DFT_adia_rep}. We then carry out a detailed comparison with existing molecular DFT approaches, namely MCDFT (Sec.~\ref{sec:diff_mcdft}), EF-DFT (Sec.~\ref{sec:EF-based_DFT}), but also electronic ensemble DFT (Sec.~\ref{sec:connections_eDFT}).} The exact Hxc density-functional energy and potentials (\ie, the first-order functional derivatives of the Hxc energy with respect to the nuclear and electronic densities, respectively), which describe the electron-electron repulsion within the molecule, are investigated further in Sec.~\ref{sec:mol_AC_formalism} by introducing a molecular adiabatic connection formalism. On that basis, a practical adiabatic density-functional approximation is proposed and discussed in Sec.~\ref{sec:mol_DFAs}. Conclusions and outlook are finally given in Sec.~\ref{sec:conclusions}. \rev{In order to make the present paper self-contained, we provide in the Appendix a detailed discussion of the BH expansion for molecular wavefunctions.}      

\section{Theory}

\rev{
\subsection{Brief review of electronic KS-DFT}\label{sec:review_elec_KSDFT}

\rev{We consider any finite system of electrons, all subject to some local and time-independent external potential $\hat{V}_{\rm ext}$ (typically created by a set of clamped nuclei in a molecule, but more general situations may be considered as well).} According to the Rayleigh--Ritz variational principle, the ground-state energy of the electronic system under study can be evaluated as follows, 
\be\label{eq:RR_VP_electronic_ener}
E^{\rm elec}_0=\min_\Phi\left\{\left\langle\hat{T}_{\rm e}+\hat{W}_{\rm ee}+\hat{V}_{\rm ext}\right\rangle_{\Phi}\right\},
\ee
where we use the shorthand notation $\langle\hat{\mathcal{O}}\rangle_\Phi=\langle \Phi\vert \hat{\mathcal{O}}\vert \Phi\rangle$ to refer to the expectation value of any quantum operator $\hat{\mathcal{O}}$, and $\Phi: r\mapsto \Phi(r)$ is a trial {\it normalized} $\mathcal{N}_\mathrm{e}$-electron wavefunction, \ie, 
\be
\langle \Phi\vert\Phi\rangle=\int dr\, \abs{\Phi(r)}^2=1, 
\ee
$r\equiv({\bf r}_1,{\bf r}_2,\ldots,{\bf r}_{\mathcal{N}_\mathrm{e}})$ denoting the set of electronic coordinates. The operators  
$\hat{T}_{\rm e}\equiv-\frac{1}{2}\sum^{\mathcal{N}_\mathrm{e}}_{i=1}\nabla_{\br_i}^2$ and $\hat{W}_{\rm ee}\equiv \sum^{\mathcal{N}_{\rm e}}_{1=i<j}\frac{1}{\abs{{\bf r}_i-{\bf r}_j}}$ (written in atomic units) describe the electronic kinetic and repulsion energies, respectively. Finally, the local external potential operator reads
\be\label{eq:ext_pot_elec_DFT}
\hat{V}_{\rm ext}\equiv \sum^{\mathcal{N}_\mathrm{e}}_{i=1}v_{\rm ext}({\bf r}_i)
\equiv \int d\br \;v_{\rm ext}({\bf r})\hat{n}(\br),  
\ee
where $\hat{n}(\br)\equiv \sum^{\mathcal{N}_\mathrm{e}}_{i=1}\delta ({\bf r}-{\bf r}_i)$ is the electron density operator at the single-electron position $\br$. According to Eq.~(\ref{eq:ext_pot_elec_DFT}), the variational principle can be simplified as follows,        
\be
E^{\rm elec}_0=\min_\Phi\left\{\left\langle\hat{T}_{\rm e}+\hat{W}_{\rm ee}\right\rangle_{\Phi}+\int d{\bf r}\;v_{\rm ext}({\bf r})n_{\Phi}(\br)\right\},
\ee
where the trial one-electron reduced density at position $\br$ (simply referred to as electronic density in the following) reads, because of the indistinguishability of the electrons,
\begin{subequations}
\begin{align}
\label{eq:elec_dens_elec_DFT}
n_{\Phi}(\br)&=\langle\hat{n}(\br)\rangle_\Phi
\\
\label{eq:elec_dens_elec_DFT_1stq}
&=\mathcal{N}_\mathrm{e}\int d{\bf r}_2\ldots \int d {\bf r}_{\mathcal{N}_\mathrm{e}} \abs{\Phi(\br,{\bf r}_2,\ldots,{\bf r}_{\mathcal{N}_\mathrm{e}})}
^2.
\end{align}
\end{subequations}
Integration over spin coordinates is implicit in the above equation (and in the rest of this work). The density emerges from the external potential energy as a natural basic variable, thus leading, through Levy's constrained-search trick~\cite{levy1979universal}, $\displaystyle\min_{\Phi}\{\ldots\}=\min_{n}\{\min_{\Phi\rightarrow n}\{\ldots\}\}$, where the minimization is performed over all wavefunctions $\Phi$ with the same density $n_\Phi(\br)=n(\br)$, to the variational principle of DFT,
\be
E^{\rm elec}_0=\min_n\left\{F[n]+\int d{\bf r}\;v_{\rm ext}({\bf r})n(\br)\right\},
\ee
where $\displaystyle F[n]=\min_{\Phi\rightarrow n}\langle\hat{T}_{\rm e}+\hat{W}_{\rm ee}\rangle_\Phi$ is the universal Levy--Lieb functional~\cite{levy1979universal,LFTransform-Lieb}. The fictitious KS non-interacting electronic system, which shares the same density $n$ as the true interacting system and is central in the standard formulation of electronic DFT (the so-called KS-DFT~\cite{Kohn1965}), finally appears when introducing the non-interacting version of $F[n]$, which reduces to a non-interacting kinetic energy functional,
\be
\displaystyle T_{\rm s}[n]=\min_{\Phi\rightarrow n}\langle\hat{T}_{\rm e}\rangle_\Phi.
\ee
On that basis, the true physical energy can be determined variationally, and in principle exactly, through the Hxc density-functional energy $E_{\rm Hxc}[n]=F[n]-T_{\rm s}[n]$:
\begin{subequations}
\begin{align}
E^{\rm elec}_0&=\min_n\left\{T_{\rm s}[n]+E_{\rm Hxc}[n]+\int d{\bf r}\;v_{\rm ext}({\bf r})n(\br)\right\}
\\
&=\min_n\min_{\Phi\rightarrow n}\left\{\left\langle\hat{T}_{\rm e}+\hat{V}_{\rm ext}\right\rangle_{\Phi}+E_{\rm Hxc}[n_\Phi]\right\}
\\
\label{eq:var_elec_ener_KSDFT}
&=
\min_\Phi \left\{\left\langle\hat{T}_{\rm e}+\hat{V}_{\rm ext}\right\rangle_{\Phi}+E_{\rm Hxc}[n_\Phi]\right\},
\end{align}
\end{subequations}
where, as readily seen from Eqs.~(\ref{eq:RR_VP_electronic_ener}) and (\ref{eq:var_elec_ener_KSDFT}), the electronic repulsion is not treated explicitly anymore but, instead, {\it via} the Hxc density functional.

\subsection{Identification of the basic variables for the molecular problem}\label{sec:identification_basic_var}

Let us now address the full molecular ($\mathcal{N}_\mathrm{e}$ electrons+ $\mathcal{N}_{\rm n}$ nuclei) problem along the lines of Sec.~\ref{sec:review_elec_KSDFT}. Our goal, ultimately reached in Sec.~\ref{sec:KS_mol_DFT}, is to map exact properties (that will be identified later in this section) onto a so-called KS molecule, by analogy with the KS electronic system, where electrons do not interact anymore among themselves. For that purpose, we start with the molecular variational principle,
\be\label{eq:GS_mol_ener_general_form_wf}
E^{\rm mol}_0=\min_\Psi\left\langle\hat{T}_{\rm n}+\hat{T}_{\rm e}+\hat{W}_{\rm ee}+\hat{V}_{\rm nn}+\hat{V}_{\rm ne}\right\rangle_{\Psi},
\ee
where $\Psi: (R,r)\mapsto \Psi(R,r)$ is a trial normalized {\it molecular} wavefunction, \ie,
\be\label{eq:normalization_mol_wf}
\langle\Psi\vert\Psi\rangle=\int dR \int dr\; \abs{\Psi(R,r)}^2=1,  
\ee
$R\equiv ({\bf R}_1,{\bf R}_2,\ldots,{\bf R}_{\mathcal{N}_\mathrm{n}})$ being the set of nuclear coordinates. The operators $\hat{T}_{\rm n}\equiv -\frac{1}{2}\sum^{\mathcal{N}_\mathrm{n}}_{i=1}\frac{\nabla_{{\mathbf{R}}_i}^2}{M_i}$,
\be 
\hat{V}_{\rm nn}\equiv {V}_{\rm nn}(R)=\sum^{\mathcal{N}_{\rm n}}_{1=i<j}\frac{\mathcal{Z}_i\mathcal{Z}_j}{\abs{{\bf R}_i-{\bf R}_j}},
\ee
and 
\be 
\hat{V}_{\rm ne}\equiv \sum^{\mathcal{N}_\mathrm{e}}_{i=1}V_{\rm ne}(R,{\bf r}_i)=-\sum^{\mathcal{N}_\mathrm{e}}_{i=1}\left(\sum^{\mathcal{N}_{\rm n}}_{j=1}\dfrac{\mathcal{Z}_j}{\abs{{\bf r}_i-{\bf R}_j}}\right)
\ee
are the nuclear kinetic energy ($M_i$ is the mass in atomic units of nucleus $i$), nuclear-nuclear repulsion ($\mathcal{Z}_i$ is the atomic number of nucleus $i$), and nuclear-electron attraction potential operators, respectively. At this point we should stress that the full molecular Hamiltonian is translationally invariant. Consequently, in order to get square integrable ground-state wavefunctions, a change of coordinates from $(R,r)$ to the total center of mass coordinates, generalized nuclear coordinates, and electronic coordinates referred to the nuclear center of mass is needed~\cite{Requist16_Exact}. For simplicity, we follow Ref.~\onlinecite{Requist16_Exact} and let $(R,r)$ denote these new coordinates. In the following, $R$ will therefore refer to a molecular geometry.\\  

For a fixed molecular geometry $R$, the nuclear-electron potential $V_{\rm ne}(R,{\bf r})$ plays the role of the external potential $v_{\rm ext}(\br)$ of electronic DFT (see Eq.~(\ref{eq:ext_pot_elec_DFT})). As the nuclei are now treated quantum mechanically, we need to reformulate the density-functional description of the electrons within the molecule. For that purpose, let us first have a closer look at the trial nuclear-electron potential energy expression, which can be simplified as follows,  
\begin{subequations}
\label{eq:Vne_expression_general_wf_form}
\begin{align}
\left\langle \hat{V}_{\rm ne}\right\rangle_{\Psi}&=\int dR \int d\br\; V_{\rm ne}(R,{\bf r})
\\
\label{eq:not_normalized_elec_dens}
&\times\mathcal{N}_{\rm e}
\int d{\bf r}_2\ldots \int d {\bf r}_{\mathcal{N}_\mathrm{e}} \abs{\Psi(R,\br,{\bf r}_2,\ldots,{\bf r}_{\mathcal{N}_\mathrm{e}})}^2
.
\end{align}
\end{subequations}
A major difference between Eqs.~(\ref{eq:not_normalized_elec_dens}) and (\ref{eq:elec_dens_elec_DFT_1stq}) is that, for a given geometry $R$, the wavefunction $\Psi(R,r)$ is {\it not} normalized with respect to the electronic coordinates, according to Eq.~(\ref{eq:normalization_mol_wf}), thus preventing us to interpret the expression in Eq.~(\ref{eq:not_normalized_elec_dens}) as an $R$-dependent electronic density and, therefore, to extend straightforwardly electronic DFT to the molecular problem. This is the reason why we now proceed with the following ``electronic normalization", for any $R$, 
\be\label{eq:normalization_for_any_R}
\Psi(R,r)\rightarrow \Phi^\Psi_{\rm eff}(R,r)=\dfrac{\Psi(R,r)}{\sqrt{\Gamma_\Psi(R)}},
\ee
where, in the normalization factor $1/\sqrt{\Gamma_\Psi(R)}$, 
\begin{subequations}
\begin{align}
\label{eq:nuclear_density_general_form_mol_wf}
&\Gamma_\Psi(R)=:\langle \Psi(R)\vert \Psi(R)\rangle_r:=\int dr\; \abs{\Psi(R,r)}^2
\\
\label{eq:nuclear_dens_intdr1dr2_etc_general_wf}
&\equiv \int d{\bf r}_1\int d{\bf r}_2\ldots \int d {\bf r}_{\mathcal{N}_\mathrm{e}}\abs{\Psi(R,\br_1,{\bf r}_2,\ldots,{\bf r}_{\mathcal{N}_\mathrm{e}})}^2 
\end{align}
\end{subequations}
is nothing but the {\it nuclear} density, and
$\Phi^\Psi_{\rm eff}(R): r\mapsto \Phi^\Psi_{\rm eff}(R,r)$ plays the role of an effective wavefunction for the electrons within the molecule in a given geometry $R$. From the ($R$-dependent) density expression at position $\br$,
\begin{subequations}
\begin{align}
\label{eq:eff_electronic_dens_general_wf}
&n^\Psi_{\rm eff}(R,\br):=n_{\Phi^\Psi_{\rm eff}(R)}(\br)=\langle \hat{n}({\bf r})\rangle_{\Phi^\Psi_{\rm eff}(R)}
\\
\label{eq:eff_electronic_dens_general_wf_1st_quantiz}
&=\mathcal{N}_\mathrm{e}\int d{\bf r}_2\ldots \int d {\bf r}_{\mathcal{N}_\mathrm{e}} \abs{\Phi^\Psi_{\rm eff}(R,\br,{\bf r}_2,\ldots,{\bf r}_{\mathcal{N}_\mathrm{e}})}
^2\\
\label{eq:eff_elec_dens_intdr2_etc_general_wf}
&=\mathcal{N}_\mathrm{e}
\dfrac{\displaystyle\int d{\bf r}_2\ldots \int d {\bf r}_{\mathcal{N}_\mathrm{e}} \abs{\Psi(R,\br,{\bf r}_2,\ldots,{\bf r}_{\mathcal{N}_\mathrm{e}})}^2}
{\displaystyle\int d{\bf r}_1\int d{\bf r}_2\ldots \int d {\bf r}_{\mathcal{N}_\mathrm{e}}\abs{\Psi(R,\br_1,{\bf r}_2,\ldots,{\bf r}_{\mathcal{N}_\mathrm{e}})}^2},
\end{align}
\end{subequations}
where we note that, by construction,
\be
\int d\br\; n^\Psi_{\rm eff}(R,\br)=\mathcal{N}_\mathrm{e},\hspace{0.2cm}\forall R,
\ee
we can finally write the trial nuclear-electron potential energy as follows, 
\be\label{eq:nuclear-electron_ener_mol_theory}
\left\langle \hat{V}_{\rm ne}\right\rangle_{\Psi}=\int dR\; \Gamma_\Psi(R) \int d\br \; V_{\rm ne}(R,{\bf r}) n^\Psi_{\rm eff}(R,\br).
\ee
The electronic wavefunction $\Phi^\Psi_{\rm eff}(R)$ (and, consequently, its density $n^\Psi_{\rm eff}(R,\br)$) is referred to as ``effective" because it does not correspond to the ground-state eigenfunction of the regular BO electronic Hamiltonian, where the nuclei are fixed in the geometry $R$. This is primarily the reason why, unlike a BO electronic solution, it can produce the exact electronic density in the presence of quantum nuclei with nuclear density $\Gamma_\Psi(R)$ at the geometry $R$, \ie,
\be
\dfrac{\delta\left\langle \hat{V}_{\rm ne}\right\rangle_{\Psi}}{\delta V_{\rm ne}(R,{\bf r})}=:\Gamma_\Psi(R)\;n^\Psi_{\rm eff}(R,\br).
\ee
Let us stress that, unlike in the EF formalism~\cite{Requist16_Exact} (see also Ref.~\onlinecite{Mayer03_book_Simple}), $\Phi^\Psi_{\rm eff}(R)$, which can be interpreted as a conditional wavefunction, will not serve as basic variable in the rest of this work. As we intend to derive a molecular KS-DFT, we will instead focus on the effective density $n^\Psi_{\rm eff}(R,\br)$, evaluated for a given molecular wavefunction $\Psi$. No assumption will be made about the mathematical form of the latter. Connections between the EF and molecular KS-DFT will be discussed later in Sec.~\ref{sec:EF-based_DFT}.    
\\

From the key expression in Eq.~(\ref{eq:nuclear-electron_ener_mol_theory}) we realize that, in order to design a KS molecule, we need to use {\it two} basic variables, unlike in electronic DFT: The geometry-dependent effective electronic density $n^\Psi_{\rm eff}(R,\br)$ {\it and} the nuclear density $\Gamma_\Psi(R)$, which encodes the quantum description of the nuclei. Note that, since the nuclear-nuclear repulsion energy is a simple and explicit functional of the nuclear density,
\begin{subequations}
\begin{align}
\left\langle \hat{V}_{\rm nn}\right\rangle_{\Psi}&= \int dR\; {V}_{\rm nn}(R)\int dr\;\abs{\Psi(R,r)}^2
\\
\label{eq:nuclear-nuclear_ener_dens_funct}
&=\int dR\; {V}_{\rm nn}(R)\,\Gamma_\Psi(R),
\end{align}
\end{subequations}
only the remaining nuclear and electronic kinetic energy contributions, as well as the electronic repulsion energy, will become implicit functionals of the electronic and nuclear densities. The derivation of a molecular KS-DFT on that basis, which is the purpose of this work, is presented in the next section.

\subsection{Molecular KS-DFT}\label{sec:KS_mol_DFT}

Following the same strategy as in electronic DFT, we start from the variational expression of the molecular ground-state energy (see Eq.~(\ref{eq:GS_mol_ener_general_form_wf})), which can be rewritten as follows, according to Eqs.~(\ref{eq:nuclear-electron_ener_mol_theory}) and (\ref{eq:nuclear-nuclear_ener_dens_funct}), 
\begin{subequations}
\begin{align}
E^{\rm mol}_0=&\min_\Psi\Bigg\{\left\langle\hat{T}_{\rm n}+\hat{T}_{\rm e}+\hat{W}_{\rm ee}\right\rangle_{\Psi}
\\
&+
\int dR\; {V}_{\rm nn}(R)\,\Gamma_\Psi(R)
\\
&+
\int dR\; \Gamma_\Psi(R) \int d\br \; V_{\rm ne}(R,{\bf r}) n^\Psi_{\rm eff}(R,\br)
\Bigg\}
.
\end{align}
\end{subequations}
If we now apply Levy's constrained search formalism~\cite{levy1979universal}, we obtain the following molecular density-functional variational principle,
\be\label{eq:mol_VP_densities}
\begin{split}
E^{\rm
mol}_0=\min_{\Gamma,n}\Big\{
&
\mathcal{F}[\Gamma,n]+\int
dR\; V_{\mathrm{nn}}(R)\,\Gamma(R)
\\
&+\int
dR\;\Gamma(R)\int d{\bf r}\;V_{\mathrm{ne}}(R,{\bf r})\,n(R,{\bf
r})\Big\}
,
\end{split}
\ee
where the analog for the molecular problem of the Levy--Lieb functional reads
\be\label{eq:mol_LL_func_general_form_wf}
\mathcal{F}[\Gamma,n]=\min_{\Psi\rightarrow
(\Gamma,n)}\left\langle
\hat{T}_{\mathrm{n}}+\hat{T}_{\mathrm{e}}+\hat{W}_{\mathrm{ee}}\right\rangle_{\Psi}.
\ee
The density constraints in the minimization of Eq.~(\ref{eq:mol_LL_func_general_form_wf}) read
\begin{subequations}
\begin{align}
\label{eq:nuclear_dens_constraint}
\Gamma_{\Psi}(R)&=\Gamma(R), 
\\
\label{eq:electron_dens_constraint}
n^\Psi_{\rm eff}(R,\br)&=n(R,{\bf
r}). 
\end{align}
\end{subequations}
}
The concept of KS molecule naturally emerges from the present molecular density-functional formalism by introducing the {\it electronically} non-interacting version of $\mathcal{F}[\Gamma,n]$,
\be\label{eq:mol_Ts_func}
\mathcal{T}_{\rm s}[\Gamma,n]=\min_{\Psi\rightarrow
(\Gamma,n)}\left\langle
\hat{T}_{\mathrm{n}}+\hat{T}_{\mathrm{e}}\right\rangle_{\Psi}.
\ee
Note that, if we can find two auxiliary (so-called KS) nuclear-nuclear $\hat{V}^{\rm KS}_{\mathrm{nn}}[\Gamma,n]$ and nuclear-electron $\hat{V}^{\rm KS}_{\mathrm{ne}}[\Gamma,n]$ potentials such that the ground state $\Psi^{\rm KS}[\Gamma,n]$
of the electronically non-interacting molecular Hamiltonian
\be
\hat{H}^{\rm KS}[\Gamma,n]=\hat{T}_{\mathrm{n}}+\hat{T}_{\mathrm{e}}+\hat{V}^{\rm KS}_{\mathrm{nn}}[\Gamma,n]+\hat{V}^{\rm KS}_{\mathrm{ne}}[\Gamma,n]
\ee
reproduces the interacting nuclear $\Gamma(R)$ and electronic $n(R,{\bf r})$ densities exactly, which will be our assumption from now on, the minimizer in Eq.~(\ref{eq:mol_Ts_func}) is $\Psi^{\rm KS}[\Gamma,n]$, since 
\be\label{eq:mol_KS_dens_func_Hamil}
\langle\hat{H}^{\rm KS}[\Gamma,n]\rangle_{\Psi^{\rm KS}[\Gamma,n]}\leq \langle\hat{H}^{\rm KS}[\Gamma,n]\rangle_{\Psi}, \hspace{0.2cm} \forall \Psi,  
\ee
according to the variational principle, and therefore,
\be
\mathcal{T}_{\rm s}[\Gamma,n]=\left\langle
\hat{T}_{\mathrm{n}}+\hat{T}_{\mathrm{e}}\right\rangle_{\Psi^{\rm KS}[\Gamma,n]}.
\ee
Complementing the above non-interacting nuclear and electronic density-functional kinetic energies with the Hxc (electronic energy within the molecule) functional
\be\label{eq:mol_Hxc_fun}
\mathcal{E}_{\rm Hxc}[\Gamma,n]=\mathcal{F}[\Gamma,n]-\mathcal{T}_{\rm s}[\Gamma,n]
\ee
restores the complete Levy--Lieb functional $\mathcal{F}[\Gamma,n]$. As discussed further in Secs.~\ref{sec:Hxc_potentials_func_deriv} and \ref{sec:mol_DFAs}, like in electronic KS-DFT, the Hxc functional will be the central quantity to approximate in order to turn the theory into a practical computational method.\\  

Returning to the exact theory, we deduce, by combining the variational principle of Eq.~(\ref{eq:mol_VP_densities}) with Eqs.~(\ref{eq:mol_Ts_func}) and (\ref{eq:mol_Hxc_fun}), the following exact ground-state molecular energy expression,
\be
\begin{split}
E^{\rm
mol}_0=\min_{\Gamma,n}\Big\{
&
\min_{\Psi\rightarrow
(\Gamma,n)}\left\langle
\hat{T}_{\mathrm{n}}+\hat{T}_{\mathrm{e}}\right\rangle_{\Psi}
\\
&+\mathcal{E}_{\rm Hxc}[\Gamma,n]+\int
dR\; V_{\mathrm{nn}}(R)\,\Gamma(R)
\\
&+\int
dR\;\Gamma(R)\int d{\bf r}\;V_{\mathrm{ne}}(R,{\bf r})\,n(R,{\bf
r})\Big\},
\end{split}
\ee
or, equivalently,
\be\label{eq:VP_KS_molecule}
\begin{split}
&E^{\rm
mol}_0=\min_{\Psi}
\Big\{\left\langle
\hat{T}_{\mathrm{n}}+\hat{T}_{\mathrm{e}}\right\rangle_{\Psi}
\\
&
+\mathcal{E}_{\rm Hxc}[\Gamma_{\Psi},n^\Psi_{\rm eff}]+\int dR\; V_{\mathrm{nn}}(R)\Gamma_{\Psi}(R)
\\
&+\int dR\;\Gamma_{\Psi}(R)\int d{\bf r}\;V_{\mathrm{ne}}(R,{\bf r})n^\Psi_{\rm eff}(R,\br)
\Big\}.
\end{split}
\ee
\rev{
The minimizer in the above equation is the KS molecular wavefunction $\Psi^{\rm KS}:=\Psi^{\rm KS}[\Gamma_0,n_0]$ that reproduces the exact nuclear $\Gamma_0(R)$ and electronic $n_0(R,\br)$ densities, \ie, those of the true physical electronically-interacting molecular wavefunction (the minimizer of Eq.~(\ref{eq:GS_mol_ener_general_form_wf})):
\begin{subequations}
\label{eq:KS_mapping_densities}
\begin{align}
\Gamma_{\Psi^{\rm KS}}(R)&=\Gamma_0(R),
\\
n^{\Psi^{\rm KS}}_{\rm eff}(R,\br):=n_{\Phi^{\Psi^{\rm KS}}_{\rm eff}(R)}(\br)&=n_0(R,\br).
\end{align}
\end{subequations}
}
According to Eq.~(\ref{eq:mol_KS_dens_func_Hamil}), it is therefore the ground-state solution to the following self-consistent electronically non-interacting KS molecular equation,
\be\label{eq:KS_mol_eq}
\left(\hat{T}_{\mathrm{n}}+\hat{T}_{\mathrm{e}}+\hat{V}^{\rm KS}_{\mathrm{nn}}+
\hat{V}^{\rm KS}_{\mathrm{ne}}\right)\Psi^{\rm KS}=\mathcal{E}^{\rm KS}\Psi^{\rm
KS},
\ee
where $\hat{V}^{\rm KS}_{\mathrm{ne}}:=\hat{V}^{\rm KS}_{\mathrm{ne}}\left[\Gamma_{\Psi^{\rm KS}},n^{\Psi^{\rm KS}}_{\rm eff}\right]$ and $\hat{V}^{\rm KS}_{\mathrm{nn}}:=\hat{V}^{\rm KS}_{\mathrm{nn}}\left[\Gamma_{\Psi^{\rm KS}},n^{\Psi^{\rm KS}}_{\rm eff}\right]$. \rev{As demonstrated in Sec.~\ref{sec:Hxc_potentials_func_deriv}, these potentials, that we refer to as nuclear-electron and nuclear-nuclear KS potentials, respectively, can be expressed as functional derivatives of the Hxc functional, \ie,
\begin{subequations}\label{eq:KS_NN_and_Ne_potentials}
\begin{align}
\label{eq:KS_Ne_potential}
V^{\rm KS}_{\mathrm{ne}}(R,{\bf r})&=V_{\mathrm{ne}}(R,{\bf r})+V^{\rm
Hxc}_{\mathrm{ne}}[\Gamma_0,n_0](R,{\bf r}),
\\
\label{eq:KS_NN_potential}
V^{\rm KS}_{\mathrm{nn}}(R)&=V_{\mathrm{nn}}(R)+V^{\rm
Hxc}_{\mathrm{nn}}[\Gamma_0,n_0](R),
\end{align}
\end{subequations}

where the nuclear-electron and nuclear-nuclear Hxc density-functional potentials read
\be\label{eq:Hxc_Ne_pot}
V^{\rm 
Hxc}_{\mathrm{ne}}[\Gamma,n](R,{\bf r})=
\dfrac{1}{\Gamma(R)}\dfrac{\delta \mathcal{E}_{\rm
Hxc}[\Gamma,n]}{\delta n(R,{\bf r})}
\ee
and
\be\label{eq:Hxc_NN_pot}
\begin{split}
V^{\rm
Hxc}_{\mathrm{nn}}[\Gamma,n](R)&=
\dfrac{\delta \mathcal{E}_{\rm
Hxc}[\Gamma,n]}{\delta
\Gamma(R)}
\\
&\quad-\dfrac{1}{\Gamma(R)}\int d{\bf r}\;\dfrac{\delta
\mathcal{E}_{\rm
Hxc}[\Gamma,n]}{\delta n(R,{\bf r})}n(R,{\bf r}),
\end{split}
\ee
respectively. Eqs.~(\ref{eq:KS_mapping_densities})--(\ref{eq:Hxc_NN_pot}) are the central result of this work. They provide an in-principle exact extension of electronic KS-DFT beyond the BO approximation. A comparison with existing molecular DFT approaches is made later in Sec.~\ref{sec:comparison}.\\
}
As a final side comment, note the division by the nuclear density $\Gamma(R)$ in the expression of the Hxc density-functional potentials (see Eqs.~(\ref{eq:Hxc_Ne_pot}) and (\ref{eq:Hxc_NN_pot})). While singularities are {\it a priori} not expected in the present ground-state theory, previous works on the EF formalism~\cite{Gossel19} suggest that instabilities may be observed when turning to the time-dependent regime. This should definitely be studied in more detail in future work.

\rev{\subsection{Born--Huang expansion of the KS molecular wavefunction}\label{sec:mol-KS_DFT_adia_rep}

\subsubsection{KS-beyond-BO electronic equations and subsequent coupled nuclear equations}
The molecular KS-DFT derived in Sec.~\ref{sec:KS_mol_DFT} is very general as no assumption is made about the mathematical form of the molecular wavefunction. For example, it could be either exactly factorized (see Sec.~\ref{sec:EF-based_DFT}) or BH-expanded. We focus on the latter choice in the following. As explained in detail in the Appendix, the (KS in the present case) molecular wavefunction can be expanded exactly {\it à la} BH~\cite{bor54}, as follows,
\be\label{eq:KS_mo_wf_BH_expansion}
\Psi^{\rm KS}(R,r)=\sum_\nu {\chi}_\nu^{\rm KS}(R)\Phi^{\rm KS}_\nu(r;R),
\ee
where the orthonormal KS electronic wavefunctions $\Phi_\nu^{\rm KS}: r\mapsto \Phi^{\rm KS}_\nu(r;R)$ vary parametrically with the geometry $R$. There is in fact a total arbitrariness in the choice of the $R$-dependent orthonormal electronic basis $\left\{\Phi^{\rm KS}_\nu\right\}$ that we refer to as KS electronic basis because it is used to describe the KS {\it molecular} wavefunction, which is defined as the solution to Eq.~(\ref{eq:KS_mol_eq}). In other words, it does not have to match the basis of ground- and excited-state KS wavefunctions that would be constructed from conventional (BO) electronic KS-DFT orbitals. We recall that the latter are optimized, in each geometry $R$, for the ground electronic state. A more natural choice in the present context consists in solving the electronic part of the molecular KS Eq.~(\ref{eq:KS_mol_eq}) for a fixed geometry $R$, by analogy with the regular adiabatic representation of (electronically-interacting) molecular wavefunctions, thus leading to  
\be\label{eq:adiabatic_rep_KS_elec_eq}
\left(\hat{T}_{\mathrm{e}}+
\sum^{\mathcal{N}_\mathrm{e}}_{i=1}{V}^{\rm KS}_{\mathrm{ne}}(R,{\bf r}_i)\right)\Phi_\nu^{\rm
KS}
={E}_\nu^{\rm KS}(R)\Phi_\nu^{\rm
KS},
\ee
where, according to Eqs.~(\ref{eq:KS_Ne_potential}) and (\ref{eq:Hxc_Ne_pot}), the KS nuclear-electron potential reads
\be\label{eq:KS_Ne_pot_used_in_adiabatic_rep}
{V}^{\rm KS}_{\mathrm{ne}}(R,{\bf r})={V}_{\mathrm{ne}}(R,{\bf r})+\dfrac{1}{\Gamma_0(R)}\dfrac{\delta \mathcal{E}_{\rm
Hxc}[\Gamma_0,n_0]}{\delta n(R,{\bf r})},
\ee
$\Gamma_0$ and $n_0$ being the exact nuclear and electronic densities (to be determined self-consistently).} This setting is convenient in practice since the solutions to Eq.~(\ref{eq:adiabatic_rep_KS_elec_eq}) are single configurational (\ie, described by a single Slater determinant or a configuration-state function). In other words, solving Eq.~(\ref{eq:adiabatic_rep_KS_elec_eq}) is equivalent to solving the following one-electron KS-like equations, 
\be\label{eq:adiabatic-like_KS_eqs}
\left(-\dfrac{\nabla^2_{\bf r}}{2}+{V}^{\rm KS}_{\mathrm{ne}}(R,{\bf r})\right)\varphi^{\rm KS}_i=\varepsilon_i(R)\varphi^{\rm KS}_i,
\ee
\rev{where $\varphi^{\rm KS}_i: \br\mapsto \varphi^{\rm KS}_i(\br;R)$ are KS-like electronic orbitals that vary parametrically with $R$. In the following, Eq.~(\ref{eq:adiabatic-like_KS_eqs}) and its solutions are referred to as ``{\it KS beyond BO}" (KS-bBO).\\
As readily seen from Eq.~(\ref{eq:KS_Ne_pot_used_in_adiabatic_rep}), Eq.~(\ref{eq:adiabatic-like_KS_eqs}) is {\it not} a regular KS equation because, unlike in standard electronic KS-DFT, it incorporates, through the Hxc functional of the nuclear and electronic densities $\mathcal{E}_{\rm
Hxc}[\Gamma,n]$, a beyond-BO description of the molecule, hence the name KS-bBO. An additional and major difference lies in the geometry-dependent electronic density $n_0(R,\br)$ that the KS-bBO orbitals are expected to reproduce, and the subsequent self-consistency loop that involves a (KS) molecular calculation (and not a purely electronic one). These two aspects are further discussed in the following. First of all, the dependence of the KS nuclear-electron potential on the nuclear wavefunctions $\left\{{\chi}_\nu^{\rm KS}\right\}$ (see Eq.~(\ref{eq:KS_mo_wf_BH_expansion})) occurs through both nuclear and electronic densities. Indeed, according to Eqs.~(\ref{eq:normalization_for_any_R}), (\ref{eq:nuclear_density_general_form_mol_wf}), (\ref{eq:eff_electronic_dens_general_wf}),
(\ref{eq:KS_mapping_densities}), and (\ref{eq:KS_mo_wf_BH_expansion}), 
\be\label{eq:nuclear_dens_from_nuclear_wfs_BH}
\Gamma_0(R)=\sum_\nu\abs{{\chi}_\nu^{\rm KS}(R)}^2,
\ee
and
\be\label{eq:eff_elec_dens_from_nuclear_and_elec_KS_wfs}
n_0(R,\br)=\dfrac{1}{\Gamma_0(R)}\sum_{\mu\nu}\left({\chi}_\mu^{\rm KS}(R)\right)^*{\chi}_\nu^{\rm KS}(R)n^{\rm KS}_{\mu\nu}(R,\br),
\ee
where we used the shorthand notation 
\be
n^{\rm KS}_{\mu\nu}(R,\br):=\langle \Phi_\mu^{\rm
KS}(R)\vert\hat{n}(\br)\vert \Phi_\nu^{\rm
KS}(R)\rangle
\ee
for the transition density elements or, equivalent, in terms of the KS-bBO orbitals and transition one-electron reduced density matrix elements (written in second quantization),  
\begin{subequations}
\label{eq:trans_density_elts_from_bBO_KS_orbs}
\begin{align}
n^{\rm KS}_{\mu\nu}(R,\br)
\equiv \sum_{ij}
&\left(\varphi^{\rm KS}_i(\br;R)\right)^*\varphi^{\rm KS}_j(\br;R)
\\
&\times\langle \Phi_\mu^{\rm
KS}\vert\hat{a}^\dagger_i\hat{a}_j\vert \Phi_\nu^{\rm
KS}\rangle
.
\end{align}
\end{subequations}
As for the molecular KS Eq.~(\ref{eq:KS_mol_eq}), projecting (and, therefore, integrating over electronic coordinates) onto the orthonormal KS-bBO states $\left\{\Phi_\nu^{\rm
KS}(R)\right\}$ leads, according to Eqs.~(\ref{eq:KS_mo_wf_BH_expansion}) and (\ref{eq:adiabatic_rep_KS_elec_eq}), to the following set of coupled nuclear equations,
\begin{subequations}
\label{eq:nuclear_eqs_mol_KS-DFT}
\begin{align}
&\Bigg(\sum_\mu\int dr\;
\left(\Phi^{\rm KS}_\nu(r;R)\right)^*\hat{T}_{\mathrm{n}}\left({\chi}_\mu^{\rm KS}(R)\Phi^{\rm KS}_\mu(r;R)\right)
\\
&+{V}^{\rm KS}_{\mathrm{nn}}(R)+{E}_\nu^{\rm KS}(R)\Bigg)\chi_\nu^{\rm
KS}(R)=\mathcal{E}^{\rm KS}\chi_\nu^{\rm
KS}(R),
\end{align}
\end{subequations}
where (see Eqs.~(\ref{eq:KS_NN_potential}) and (\ref{eq:Hxc_NN_pot}))
\begin{subequations}
\label{eq:KS_nn_pot_complete_expression_adia_calculations}
\begin{align}
{V}^{\rm KS}_{\mathrm{nn}}(R)&=V_{\mathrm{nn}}(R)+\dfrac{\delta \mathcal{E}_{\rm
Hxc}[\Gamma_0,n_0]}{\delta
\Gamma(R)}
\\
&\quad 
-\dfrac{1}{\Gamma_0(R)}\int d{\bf r}\;\dfrac{\delta
\mathcal{E}_{\rm
Hxc}[\Gamma_0,n_0]}{\delta n(R,{\bf r})}n_0(R,{\bf r}),
\end{align}
\end{subequations}
and the non-adiabatic couplings between states $\nu$ and $\mu$ originate from the nuclear kinetic energy operator $\hat{T}_{\mathrm{n}}$ and the $R$-dependence of the KS-bBO states (see the Appendix for further details).\\
}
As a final comment, note that the molecular KS wavefunction form of Eq.~(\ref{eq:KS_mo_wf_BH_expansion}) is invariant \rev{under} unitary
transformations applied to both nuclear and $R$-dependent electronic wavefunctions. Thus, exploring the alternative option of generating a (quasi)diabatic many-body KS representation in this context -- helped by a gauge-invariant unitary transformation constraint for choosing ``least-changing-with-$R$'' (diabatic, non-canonical KS, yet still natural) orbitals and corresponding single-configuration KS electronic wavefunctions -- is a tempting avenue (for neglecting nonadiabatic couplings induced by $\hat{T}_{\mathrm{n}}$), which will obviously deserve further investigation in future work.

\rev{
\subsubsection{Self-consistency at the full KS molecular level}

 Let us now have a closer look at the self-consistent evaluation of both electronic and nuclear solutions to the coupled Eqs.~(\ref{eq:adiabatic_rep_KS_elec_eq}) [or, equivalently, Eq.~(\ref{eq:adiabatic-like_KS_eqs})] and (\ref{eq:nuclear_eqs_mol_KS-DFT}), respectively. For trial sets $\left\{{\chi}_\nu^{{\rm KS}(0)}\right\}$ and $\left\{\varphi^{{\rm KS}(0)}_i\right\}$ of nuclear wavefunctions and KS-bBO orbitals (one could use the ground-state vibrational wavefunction and conventional BO KS orbitals as a guess, for example), a trial nuclear density $\Gamma^{(0)}_0(R)$  can be evaluated, according to Eq.~(\ref{eq:nuclear_dens_from_nuclear_wfs_BH}), as well as a trial electronic density $n^{(0)}_0(R,\br)$, according to Eq.~(\ref{eq:trans_density_elts_from_bBO_KS_orbs}), thus generating a trial nuclear-electron KS potential ${V}^{{\rm KS}(0)}_{\mathrm{ne}}(R,{\bf r})$, according to Eq.~(\ref{eq:KS_Ne_pot_used_in_adiabatic_rep}). Consequently, the KS-bBO Eq.~(\ref{eq:adiabatic-like_KS_eqs}) can be solved, thus generating an updated set $\left\{\varphi^{{\rm KS}(1)}_i\right\}$ of orbitals (and, therefore, an updated set $\left\{\Phi_\nu^{{\rm
KS}(1)}\right\}$ of $R$-dependent KS-bBO electronic wavefunctions) as well as an updated electronic density $n^{(1)}_0(R,\br)$ from which, in conjunction with $\Gamma^{(0)}_0(R)$ (see Eq.~(\ref{eq:KS_nn_pot_complete_expression_adia_calculations})), the nuclear Eqs.~(\ref{eq:nuclear_eqs_mol_KS-DFT}) can be solved. Thus we can finally generate a new set $\left\{{\chi}_\nu^{{\rm KS}(1)}\right\}$ of nuclear wavefunctions. A this point, we have updated both nuclear wavefunctions and KS-bBO orbitals, which enables to start a new self-consistency loop, until convergence is reached.\\
The above-mentioned self-consistent implementation of molecular KS-DFT, which relies on the BH expansion of the KS molecular wavefunction (see Eq.~(\ref{eq:KS_mo_wf_BH_expansion})), is probably the most appealing formulation of the theory from a practical point of view. In particular, in this context, evaluating the non-adiabatic couplings (see the Appendix) directly from the KS-bBO orbitals is not an approximation, unlike in conventional beyond-BO simulations based on BO KS-DFT computations, where response TDDFT is required~\cite{Wang21_NAC-TDDFT,Send10_First-order,Ou15_First-order}. This drastic simplification originates from the fact that electrons are non-interacting in the (auxiliary) KS molecule from which, it should be remembered, the exact ground-state energy of the true electronically-interacting molecule can be recovered (see Eq.~(\ref{eq:VP_KS_molecule})). From the perspective of EF-DFT~\cite{Requist16_Exact,Li18_Density}, to which molecular KS-DFT is compared later in Sec.~\ref{sec:EF-based_DFT}, we may provide within such a scheme a simpler (because more explicit, through the BH expansion) dependence of the Hxc energy on both the paramagnetic current density and the quantum geometric tensor. This should be further explored in future work.   
} 
 
\rev{

\section{Comparison with existing density-functional approaches}\label{sec:comparison}

\subsection{Comparison with multi-component DFT}\label{sec:diff_mcdft}

Even though molecular KS-DFT and MCDFT~\cite{Kreibich2001,Gidopoulos98,Butriy07,Kreibich08} (see also Refs.~\onlinecite{Dreizler90_Density,Capitani82_Non‐Born}) have a basic variable in common, namely the nuclear density $\Gamma$, they differ substantially in two ways. First of all, in MCDFT, the one-electron density associated to a molecular wavefunction $\Psi$ is defined differently, as follows,    
\be\label{eq:one-el_dens_MCDFT}
n_\Psi(\br_1)\overset{\textrm{MCDFT}}{:=}\mathcal{N}_e\int d\br_2\ldots \int d\br_{\mathcal{N}_e} \int dR\, \abs{\Psi(R,r)}^2,
\ee
where we recall that $\Psi(R,r)\equiv\Psi(R,\br_1,\br_2,\ldots,\br_{\mathcal{N}_e})$. As readily seen from Eq.~(\ref{eq:one-el_dens_MCDFT}), unlike the effective electronic density $n^\Psi_{\rm eff}(R,\br)$ introduced in Eq.~(\ref{eq:eff_electronic_dens_general_wf_1st_quantiz}), which is the additional basic variable in molecular KS-DFT, $n_\Psi$ is geometry-{\it independent}. Consequently, the nuclear-electron potential energy cannot be expressed explicitly as a functional of $\Gamma_\Psi$ and $n_\Psi$ (see Eq.~(\ref{eq:Vne_expression_general_wf_form})). In MCDFT, the simplification of the molecular problem originates from the fact that the two densities are mapped onto a fictitious molecule where both the electronic repulsion {\it and} the nuclear-electron attraction are turned off. On the other hand, in molecular KS-DFT, the nuclear-electron potential energy is expressed as an explicit functional of $\Gamma_\Psi$ and $n^\Psi_{\rm eff}$, so that, in the KS molecule, nuclei and electrons still interact, like in standard beyond-BO simulations based on BO KS-DFT and linear response TDDFT computations. While molecular KS-DFT only achieves a density-functional simplification of the electronic structure problem within the molecule, further simplifications can be envisaged in this context through an EF~\cite{Abedi10_EF,Villaseco22_Exact} of the molecular KS wavefunction. This is further discussed in Sec.~\ref{sec:EF-based_DFT}. 

\subsection{Comparison with the exact-factorization-based molecular DFT}\label{sec:EF-based_DFT}

As already pointed out in Sec.~\ref{sec:mol-KS_DFT_adia_rep}, no assumption is made, in the derivation of molecular KS-DFT (see Sec.~\ref{sec:KS_mol_DFT}), about the mathematical form of the molecular wavefunction. If we employ the following exactly-factorized expression~\cite{Requist16_Exact} (which has been originally introduced as a way to exactify the BO approximation~\cite{Villaseco22_Exact}) for a trial molecular wavefunction $\Psi$,
\be\label{eq:EF_mol_wf}
\Psi(R,r)=\chi(R)\Phi_R(r),
\ee
where the so-called conditional electronic wavefunction $\Phi_R:r\mapsto \Phi_R(r)$ is normalized, for any $R$, \ie, 
\begin{subequations}
\begin{align}
&\int dr\; \vert\Phi_R(r)\vert^2
\\
&\equiv \int d{\bf r}_1\int d{\bf r}_2\ldots \int d {\bf r}_{\mathcal{N}_\mathrm{e}}\abs{\Phi_R(\br_1,{\bf r}_2,\ldots,{\bf r}_{\mathcal{N}_\mathrm{e}})}^2 
\\
&=1,\hspace{0.2cm} \forall R,
\end{align}
\end{subequations}
and $\chi(R)$ is the marginal (nuclear) wavefunction, our two basic variables (the nuclear and effective electronic densities) read as follows, according to Eqs.~(\ref{eq:nuclear_dens_intdr1dr2_etc_general_wf}) and (\ref{eq:eff_elec_dens_intdr2_etc_general_wf}),
\be
\Gamma_\Psi(R)=\abs{\chi(R)}^2,
\ee
and
\begin{subequations}
\begin{align}
n^\Psi_{\rm eff}(R,\br)&=\mathcal{N}_{\rm e}\int d{\bf r}_2\ldots \int d {\bf r}_{\mathcal{N}_\mathrm{e}}\abs{\Phi_R(\br,{\bf r}_2,\ldots,{\bf r}_{\mathcal{N}_\mathrm{e}})}^2
\\
&\equiv n_{\Phi_R}(\br). 
\end{align}
\end{subequations}
Therefore, the conditional density of the EF matches the effective electronic density of molecular KS-DFT.\\

Obviously, the EF formalism can be merged with molecular KS-DFT, simply by considering an exactly factorized form of the KS molecular wavefunction, 
\be
\Psi^{\rm KS}(R,r)=\chi^{\rm KS}(R)\Phi^{\rm KS}_R(r),
\ee
where the marginal and conditional KS wavefunctions reproduce the exact nuclear and effective electronic densities, respectively (see Eq.~(\ref{eq:KS_mapping_densities})), \ie, 
\be
\abs{\chi^{\rm KS}(R)}^2=\Gamma_0(R),
\ee
and
\be
n_{\Phi^{\rm KS}_R}(\br)=n_0(R,\br).
\ee
Even though the EF-DFT of Requist and Gross~\cite{Requist16_Exact} reproduces the two above-mentioned densities, it still differs from the present molecular KS-DFT. First of all, in EF-DFT, the true physical marginal wavefunction (evaluated for interacting electrons) is determined, while $\chi^{\rm KS}(R)$ describes nuclei that interact (within the KS molecule) with non-interacting electrons. Secondly, the KS conditional wavefunction in EF-DFT is expected to reproduce not only the effective electronic density in all geometries, but also the paramagnetic current density of the true physical conditional wavefunction. The latter constraint is not imposed in molecular KS-DFT. Still, the theory is equivalently in-principle exact because the exact Hxc functional $\mathcal{E}_{\rm Hxc}\left[\Gamma,n\right]$ is not local geometry-wise. This point is further discussed in Sec.~\ref{sec:mol_DFAs}. The use of the EF within molecular KS-DFT should be further explored as well as its potential practical advantage over the formulation {\it à la} BH given in Sec.~\ref{sec:mol-KS_DFT_adia_rep}.       

\subsection{Comparison with electronic ensemble DFT}\label{sec:connections_eDFT}

A major difference between conventional KS-DFT equations and the KS-bBO ones (see Eq.~(\ref{eq:adiabatic-like_KS_eqs})) is that, in the latter case, both ground- and excited-state electronic configurations contribute to the electronic density and, therefore, to the total molecular energy. Indeed, by rewriting the latter as follows (see Eqs.~(\ref{eq:VP_KS_molecule}) and (\ref{eq:KS_Ne_potential})),
\begin{subequations}
\begin{align}
E_0^{\rm mol}=&\langle\hat{T}_{\rm n}\rangle_{\Psi^{\rm KS}}+\mathcal{E}_{\rm Hxc}\left[\Gamma_0,n_0\right]
\\
&+\int dR\;V_{\mathrm{nn}}(R)\,\Gamma_0(R)
\\
&-\int dR\;\Gamma_0(R)\int d{\bf r}\;V^{\rm Hxc}_{\mathrm{ne}}(R,{\bf r})n_{0}(R,{\bf
r})
\\
\label{eq:T_e_plus_VKS_ne}
&+\langle\hat{T}_{\rm e}+\hat{V}^{\rm KS}_{\rm ne}\rangle_{\Psi^{\rm KS}},
\end{align}
\end{subequations}
where the physical nuclear-electron potential has simply been replaced by the difference between its KS and Hxc analogues, and inserting the BH expansion of the KS molecular wavefunction (see Eq.~(\ref{eq:KS_mo_wf_BH_expansion})) into Eq.~(\ref{eq:T_e_plus_VKS_ne}), we obtain the following expression, according to the many-electron KS-bBO Eq.~(\ref{eq:adiabatic_rep_KS_elec_eq}) and Eq.~(\ref{eq:nuclear_dens_from_nuclear_wfs_BH}),  
\begin{subequations}
\begin{align}
&E_0^{\rm mol}=\langle\hat{T}_{\rm n}\rangle_{\Psi^{\rm KS}}+\mathcal{E}_{\rm Hxc}\left[\sum_\nu \vert\chi^{\rm KS}_\nu \vert^2,n_0\right]
\\
&-\int dR\;\sum_\nu \vert\chi^{\rm KS}_\nu(R) \vert^2\int d{\bf r}\;V^{\rm Hxc}_{\mathrm{ne}}(R,{\bf r})n_{0}(R,{\bf
r})
\\
\label{eq:ens_KS_elec_ener_in_E0mol}
&+\int dR\sum_\nu \vert\chi^{\rm KS}_\nu(R) \vert^2\left({V}_{\mathrm{nn}}(R)+E^{{\rm KS},\underline{\xi}(R)}(R)\right).
\end{align}
\end{subequations}
The KS electronic energy contribution (second term in Eq.~(\ref{eq:ens_KS_elec_ener_in_E0mol})) turns out to be an ensemble KS energy~\cite{JPC79_Theophilou_equi-ensembles,theophilou_book,gross1988rayleigh,gross1988density,oliveira1988density,Cernatic2022,Gould2023_Electronic,scott2023exact},  
\be
E^{{\rm KS},\underline{\xi}(R)}(R)=\sum_\nu \xi_\nu(R)\;{E}_\nu^{\rm KS}(R),
\ee
where the geometry-dependent set $\underline{\xi}(R)$ of ensemble weights is determined from the nuclear wavefunctions as follows, 
\be
\underline{\xi}(R)\equiv \left\{\xi_\nu(R)=\dfrac{\vert\chi^{\rm KS}_\nu(R) \vert^2}{\sum_\mu \vert\chi^{\rm KS}_\mu(R) \vert^2}=\dfrac{\vert\chi^{\rm KS}_\nu(R) \vert^2}{\Gamma_0(R)}\right\}.
\ee
Despite this similarity with ensemble DFT, the KS-bBO DFT description of the electrons within the molecule is not an ensemble density-functional one, simply because the corresponding ensemble density,
\begin{subequations}
\begin{align}
n^{\underline{\xi}(R)}(R,{\bf r})&:=\sum_\nu \xi_\nu(R)\;n_{\Phi^{\rm KS}_\nu(R)}({\bf r})
\\
&=\dfrac{1}{\Gamma_0(R)}\sum_{\nu}\vert\chi^{\rm KS}_\nu(R) \vert^2 n^{\rm KS}_{\nu\nu}(R,\br),
\end{align}
\end{subequations}
misses the off-diagonal $\mu\neq\nu$ contributions to the effective electronic density that we map onto the KS molecule (see Eq.~(\ref{eq:eff_elec_dens_from_nuclear_and_elec_KS_wfs})). Moreover, the KS-bBO (nuclear-electron) potential (see Eq.~(\ref{eq:KS_Ne_pot_used_in_adiabatic_rep})) is a functional of the nuclear density $\sum_\mu \vert\chi^{\rm KS}_\mu(R) \vert^2$ while an ensemble KS potential~\cite{Cernatic2022} would be, in a given geometry $R$, a function of the ensemble weights $\left\{\xi_\nu(R)=\vert\chi^{\rm KS}_\nu(R)\vert^2/\sum_\mu \vert\chi^{\rm KS}_\mu(R) \vert^2\right\}$.\\
In summary, even though formulating molecular KS-DFT {\it à la} BH involves both ground- and excited-state electronic configurations, the resulting description of the electronic structure within the molecule deviates substantially from that of a density-functional KS ensemble.   

}

\section{Molecular adiabatic connection formalism}\label{sec:mol_AC_formalism}

In this section we will prove the expressions given in Eqs.~(\ref{eq:KS_NN_and_Ne_potentials})--(\ref{eq:Hxc_NN_pot}) for the nuclear-nuclear and nuclear-electron KS density-functional potentials. Moreover, as readily seen from Eqs.~(\ref{eq:mol_LL_func_general_form_wf}), (\ref{eq:mol_Ts_func}), and (\ref{eq:mol_Hxc_fun}), the Hxc density functional $\mathcal{E}_{\rm Hxc}[\Gamma,n]$ describes not only the electronic repulsion energy but also nuclear and electronic kinetic energy corrections induced by the electronic repulsion. Establishing a clearer connection between the former and the latter, by analogy with electronic KS-DFT, is highly desirable, in particular for the development of density-functional approximations in this context.

\subsection{Definition of the adiabatic connection path}

 We propose to adapt the coupling-constant integration technique of electronic DFT~\cite{Harris1974,LANGRETH1975,Gunnarsson76,Langreth77_Exchange-correlation}, known as the adiabatic connection (AC) formalism, to the full molecular problem. For that purpose, we consider the following molecular Schr\"{o}dinger equation, with the partial electronic repulsion operator $\lambda\hat{W}_{\mathrm{ee}}$,      
\be\label{eq:AC_SE}
\left(\hat{T}_{\mathrm{n}}+\hat{T}_{\mathrm{e}}+\lambda\hat{W}_{\mathrm{ee}}+\hat{V}^\lambda_{\mathrm{nn}}+\hat{V}^\lambda_{\mathrm{ne}}\right)\Psi^\lambda=\mathcal{E}^\lambda\Psi^\lambda,
\ee
where $\lambda$ varies in the range $0\leq \lambda\leq 1$, and $\Psi^\lambda$ is the {\it normalized} partially electronically-interacting ground-state solution. 
Note that, even though it is not compulsory (and not used in the following derivations), we could, for implementing such an AC, use the adiabatic electronic representation (see Sec.~\ref{sec:mol-KS_DFT_adia_rep}) based on the following $R$-dependent and partially-interacting electronic Schr\"{o}dinger equation, 
\be\label{eq:adiabatic_rep_AC}
\left(\hat{T}_{\mathrm{e}}+\lambda\hat{W}_{\mathrm{ee}}+\sum^{\mathcal{N}_e}_{i=1}{V}^\lambda_{\mathrm{ne}}(R,{\bf
r}_i)\right)\Phi^\lambda_\nu=E^\lambda_\nu(R)\Phi^\lambda_\nu,
\ee
where the ground- and excited-state solutions $\Phi^\lambda_\nu: r\mapsto \Phi^\lambda_\nu(r;R)$ vary parametrically with the geometry $R$, thus leading to the following BH expansion of the full molecular wavefunction,
\be\label{eq:BH_expansion_AC_lambda-int}
\Psi^\lambda(R,r)=\sum_\nu {\chi}_\nu^\lambda(R)\Phi^\lambda_\nu(r;R). 
\ee
The above expression will become useful later on (in Sec.~\ref{sec:mol_DFAs}) for assessing density-functional approximations in the present context.\\ 

The key ingredients in the present AC are the $\lambda$-dependent nuclear-nuclear ${V}^\lambda_{\mathrm{nn}}(R)$ and nuclear-electron ${V}^\lambda_{\mathrm{ne}}(R,{\bf r})$ potentials that are adjusted such that the following density constraints
are fulfilled for {\it any} interaction strength $\lambda$,
\be
\label{eq:AC_Gamma_constraint}
\Gamma_{\Psi^\lambda}(R)&=&\Gamma(R), \hspace{0.2cm}\forall \lambda\in [0, 1],
\ee
and
\be
\label{eq:AC_elecdens_constraint}
\begin{split}
n_{\Phi^\lambda_{\rm eff}(R)}({\bf r})&:=n^{\Psi^{\lambda}}_{\rm eff}(R,{\bf
r})
\\
&=n(R,{\bf
r}), \hspace{0.2cm}\forall \lambda\in [0, 1],
\end{split}
\ee
where we used the shorthand notation (see Eqs.~(\ref{eq:normalization_for_any_R}) and (\ref{eq:AC_Gamma_constraint}))
\be\label{eq:def_Phi-lambda_eff}
\Phi^\lambda_{\rm eff}(R):r\mapsto \Phi^\lambda_{\rm eff}(R,r)=\Phi^{\Psi^\lambda}_{\rm eff}(R,r)=\dfrac{\Psi^\lambda(R,r)}{\sqrt{\Gamma(R)}}.
\ee
Consequently, for a given $\lambda$ value, these potentials as well as the partially-interacting ground molecular state become functionals of the nuclear and electronic densities:
\be
\begin{split}
\Psi^\lambda(R,r)&\equiv\Psi^\lambda[\Gamma,n](R,r),\\
V^\lambda_{\mathrm{nn}}(R)&\equiv V^\lambda_{\mathrm{nn}}[\Gamma,n](R),
\\
V^\lambda_{\mathrm{ne}}(R,{\bf
r})&\equiv V^\lambda_{\mathrm{ne}}[\Gamma,n](R,{\bf r}).
\end{split}
\ee
Thus, we can introduce the following partially-interacting Levy--Lieb functional, 
\be\label{eq:mol_LL_functional}
\mathcal{F}^\lambda[\Gamma,n]&=&\min_{\Psi\rightarrow
(\Gamma,n)}\left\langle
\hat{T}_{\mathrm{n}}+\hat{T}_{\mathrm{e}}+\lambda\hat{W}_{\mathrm{ee}}\right\rangle_{\Psi}
\\
&=&
\left\langle
\hat{T}_{\mathrm{n}}+\hat{T}_{\mathrm{e}}+\lambda\hat{W}_{\mathrm{ee}}\right\rangle_{\Psi^\lambda},
\ee
from which the molecular Levy--Lieb $\mathcal{F}[\Gamma,n]=\mathcal{F}^{\lambda=1}[\Gamma,n]$ and non-interacting kinetic energy $\mathcal{T}_{\rm s}[\Gamma,n]=\mathcal{F}^{\lambda=0}[\Gamma,n]$ functionals are recovered in the fully ($\lambda=1$) and non-interacting ($\lambda=0$) limits, respectively.\\ 

\subsection{KS potentials as functional derivatives}\label{sec:Hxc_potentials_func_deriv}

From Eq.~(\ref{eq:AC_SE}), the density constraints of Eqs.~(\ref{eq:AC_Gamma_constraint}) and (\ref{eq:AC_elecdens_constraint}), and the normalization of $\Psi^\lambda$ for any choice of $\Gamma$ and $n$, it comes
\be
\begin{split}
&\dfrac{\delta\mathcal{F}^\lambda[\Gamma,n]}{\delta n(R,{\bf r})}=
\left\langle\frac{\delta\Psi^\lambda}{\delta n(R,{\bf r})}\middle\vert\hat{T}_{\mathrm{n}}+\hat{T}_{\mathrm{e}}+\lambda\hat{W}_{\mathrm{ee}}\middle\vert\Psi^\lambda\right\rangle+c.c.
\\
&=-\left\langle\frac{\delta\Psi^\lambda}{\delta n(R,{\bf r})}\middle\vert\hat{V}^\lambda_{\mathrm{ne}}+\hat{V}^\lambda_{\mathrm{nn}}\middle\vert\Psi^\lambda\right\rangle-c.c.
\\
&=-\left.\dfrac{\delta \left[\int dR\int d{\bf r}\;\Gamma(R)\mathcal{V}_{\mathrm{ne}}(R,{\bf r}) n(R,{\bf r})\right]}{\delta n(R,{\bf r})}\right|_{\mathcal{V}_{\mathrm{ne}}=V^\lambda_{\mathrm{ne}}},
%
\end{split}
\ee
thus leading to 
\be\label{eq:lambda_Ne_pot_dens_func_deriv}
V^\lambda_{\mathrm{ne}}(R,{\bf r})=-\dfrac{1}{\Gamma(R)}\dfrac{\delta
\mathcal{F}^\lambda[\Gamma,n]}{\delta n(R,{\bf r})}.
\ee
Similarly, 
\be
\begin{split}
&\dfrac{\delta\mathcal{F}^\lambda[\Gamma,n]}{\delta \Gamma(R)}=
\left\langle\frac{\delta\Psi^\lambda}{\delta \Gamma(R)}\middle\vert\hat{T}_{\mathrm{n}}+\hat{T}_{\mathrm{e}}+\lambda\hat{W}_{\mathrm{ee}}\middle\vert\Psi^\lambda\right\rangle+c.c.
\\
&=-\left\langle\frac{\delta\Psi^\lambda}{\delta \Gamma(R)}\middle\vert\hat{V}^\lambda_{\mathrm{ne}}+\hat{V}^\lambda_{\mathrm{nn}}\middle\vert\Psi^\lambda\right\rangle-c.c.
\\
&=-\left.\dfrac{\delta \left[\int dR\;\Gamma(R)\int d{\bf r}\,\mathcal{V}_{\mathrm{ne}}(R,{\bf r}) n(R,{\bf r})\right]}{\delta \Gamma(R)}\right|_{\mathcal{V}_{\mathrm{ne}}=V^\lambda_{\mathrm{ne}}}
\\
&\quad\quad-\left.\dfrac{\delta \left[\int dR\;\mathcal{V}_{\mathrm{nn}}(R)\Gamma(R)\right]}{\delta \Gamma(R)}\right|_{\mathcal{V}_{\mathrm{nn}}=V^\lambda_{\mathrm{nn}}},
\end{split}
\ee
where both functional derivatives are evaluated for {\it fixed} nuclear-electron and nuclear-nuclear potentials,
thus leading to
\be
\begin{split}
V^\lambda_{\mathrm{nn}}(R)=-\dfrac{\delta \mathcal{F}^\lambda[\Gamma,n]}{\delta
\Gamma(R)}
-\int d{\bf r}\;V^\lambda_{\mathrm{ne}}(R,{\bf r}) n(R,{\bf r}),
\end{split}
\ee
or, equivalently, according to Eq.~(\ref{eq:lambda_Ne_pot_dens_func_deriv}),
\be
\begin{split}
V^\lambda_{\mathrm{nn}}(R)&=-\dfrac{\delta \mathcal{F}^\lambda[\Gamma,n]}{\delta
\Gamma(R)}
\\
&\quad+\dfrac{1}{\Gamma(R)}\int d{\bf r}\;\dfrac{\delta
\mathcal{F}^\lambda[\Gamma,n]}{\delta n(R,{\bf r})}n(R,{\bf r}).
\end{split}
\ee
By considering the particular case where the densities are equal to the true physical ones $\Gamma_0$ and $n_0$ of the molecule under study, \rev{and rewriting the nuclear-electron and nuclear-nuclear KS potentials as follows,
\begin{subequations}
\begin{align}
V^{\rm KS}_{\mathrm{ne}}(R,{\bf r})&=V^{\lambda=1}_{\mathrm{ne}}(R,{\bf
r})+\left(V^{\lambda=0}_{\mathrm{ne}}(R,{\bf
r})-V^{\lambda=1}_{\mathrm{ne}}(R,{\bf
r})\right),
\\
V^{\rm KS}_{\mathrm{nn}}(R)
&=V^{\lambda=1}_{\mathrm{nn}}(R,{\bf
r})+\left(V^{\lambda=0}_{\mathrm{nn}}(R,{\bf
r})-V^{\lambda=1}_{\mathrm{nn}}(R,{\bf
r})\right),
\end{align}
\end{subequations}
respectively, we finally prove, simply by referring to the definition of the Hxc functional in Eq.~(\ref{eq:mol_Hxc_fun}), the exact expressions given in Eq.~(\ref{eq:KS_NN_and_Ne_potentials}).} 

\subsection{Adiabatic connection formula for the Hxc energy within the molecule}

Unlike in electronic KS-DFT, the Hxc functional $\mathcal{E}_{\rm
Hxc}[\Gamma,n]$ introduced in Eq.~(\ref{eq:mol_Hxc_fun}) is not universal \rev{in the sense that} it depends on the nuclear kinetic energy operator $\hat{T}_{\mathrm{n}}$ (see Eqs.~(\ref{eq:mol_LL_func_general_form_wf}) and (\ref{eq:mol_Ts_func})), which
is system-dependent \rev{(in particular because it varies with the masses of the nuclei)}. However, it depends neither on the nuclear-nuclear potential nor on the nuclear-electron one. It is universal in this respect. By analogy with electronic KS-DFT, it is possible to make the dependence on $\hat{T}_{\mathrm{n}}$ implicit, which, as we will see in Sec.~\ref{sec:mol_DFAs}, can be exploited in the design of density-functional approximations.\\

The exact Hxc functional can be rewritten as follows (see Eq.~(\ref{eq:mol_Hxc_fun})),
\be\label{eq:mol_Hxc_int_lambda_dFlambda_dlambda}
\mathcal{E}_{\rm
Hxc}[\Gamma,n]=\int^1_0 d\lambda \dfrac{d\mathcal{F}^\lambda[\Gamma,n]}{d\lambda},
\ee
where, according to Eqs.~(\ref{eq:mol_LL_functional}) and (\ref{eq:AC_SE}), and
the density constraints of Eqs.~(\ref{eq:AC_Gamma_constraint}) and (\ref{eq:AC_elecdens_constraint}),   
\be
\begin{split}
\mathcal{F}^\lambda[\Gamma,n]&=\mathcal{E}^\lambda-\int dR\;\Gamma(R)V^\lambda_{\mathrm{nn}}(R)
\\
&\quad-\int
dR\;\Gamma(R)\int d{\bf r}\,V^\lambda_{\mathrm{ne}}(R,{\bf r})n(R,{\bf r}).
\end{split}
\ee  
Since, according to the Hellmann--Feynman theorem,
\be
\begin{split}
\dfrac{d\mathcal{F}^\lambda[\Gamma,n]}{d\lambda}&=\left\langle
\hat{W}_{\mathrm{ee}}+\dfrac{d\hat{V}^\lambda_{\mathrm{nn}}}{d\lambda}+\dfrac{d\hat{V}^\lambda_{\mathrm{ne}}}{d\lambda}\right\rangle_{\Psi^\lambda}
\\
&\quad-\int dR\;\Gamma(R)\dfrac{\partial V^\lambda_{\mathrm{nn}}(R)}{\partial \lambda}
\\
&\quad-\int
dR\;\Gamma(R)\int d{\bf r}\,\dfrac{\partial V^\lambda_{\mathrm{ne}}(R,{\bf
r})}{\partial \lambda}n(R,{\bf r}),
\end{split}
\ee
or, equivalently,
\be
\dfrac{d\mathcal{F}^\lambda[\Gamma,n]}{d\lambda}=\left\langle
\hat{W}_{\mathrm{ee}}\right\rangle_{\Psi^\lambda},
\ee
thus leading to (with the notation of Eq.~(\ref{eq:def_Phi-lambda_eff}))
\be
\dfrac{d\mathcal{F}^\lambda[\Gamma,n]}{d\lambda}=
\int dR\;\Gamma(R)\left\langle
\hat{W}_{\mathrm{ee}}\right\rangle_{\Phi^\lambda_{\rm eff}(R)},
\ee
we deduce from Eq.~(\ref{eq:mol_Hxc_int_lambda_dFlambda_dlambda}) the following exact AC-based expression:
\be\label{eq:mol_AC_formula_Hxc_ener}
\mathcal{E}_{\rm
Hxc}[\Gamma,n]=\int dR\;\Gamma(R)
\int^1_0 d\lambda
\left\langle
\hat{W}_{\mathrm{ee}}\right\rangle_{\Phi^\lambda_{\rm eff}(R)}.
\ee

\rev{ 
\section{Practical adiabatic density-functional approximation}\label{sec:mol_DFAs}
}
In electronic KS-DFT, the exact Hxc density-functional energy can be
expressed as follows~\cite{Langreth77_Exchange-correlation},
\be
E_{\rm Hxc}[\rho]=\int^1_0 d\lambda
\left\langle
\hat{W}_{\mathrm{ee}}\right\rangle_{\Phi^\lambda[\rho]},
\ee
where $\Phi^\lambda[\rho]$ is the ground state of the partially-interacting
electronic Hamiltonian
$\hat{T}_{\mathrm{e}}+\lambda\hat{W}_{\mathrm{ee}}+\sum^{\mathcal{N}_{\mathrm{e}}}_{i=1}v^{\lambda}[\rho]({\bf
r}_i)$ with density $n_{\Phi^\lambda[\rho]}({\bf r})=\rho({\bf r})$,
$\forall \lambda\in[0,1]$. Note that the electronic density $\rho({\bf r})$
has no geometry dependence, unlike the effective electronic density $n(R,{\bf
r})$ of molecular KS-DFT. Let us also stress that, unlike
$\Phi^\lambda[\rho](r)$, which is a pure partially-interacting electronic
ground state, the partially-interacting effective electronic wavefunction of the molecular AC, $\Phi^\lambda_{\rm
eff}(R,r)=\sum_\nu\chi^\lambda_\nu(R)\Phi^\lambda_\nu(r;R)/\sqrt{\Gamma(R)}$,
is a linear combination of partially-interacting ground and excited
electronic states (see Eqs.~(\ref{eq:adiabatic_rep_AC}), (\ref{eq:BH_expansion_AC_lambda-int}), and (\ref{eq:def_Phi-lambda_eff})). It has {\it a priori} no reason to be itself, in a given geometry $R$, the ground state of a partially-interacting electronic Hamiltonian. Moreover, $\Phi^\lambda_{\rm eff}\equiv \Phi^\lambda_{\rm eff}[\Gamma, n]$ is a functional of the electronic density $n$ within the molecule, which means that it is in principle determined from the values that $n(R,{\bf r})$ can take for various geometries $R$. This can be related to the fact that the non-adiabatic couplings (\ie, the derivatives of all electronic wavefunctions $\left\{{\Phi}_\nu^\lambda(r;R)\right\}$ with respect to $R$) are needed to determine the nuclear wavefunctions $\left\{{\chi}_\nu^\lambda(R)\right\}$ and, therefore, $\Phi^\lambda_{\rm eff}(R,r)$. Consequently, for the above reasons, the latter is not expected to match its adiabatic approximation \rev{$\Phi^\lambda[n_R](r)$, where $n_R: \br\mapsto n_R(\br)=n(R,{\bf r})$ is the effective electronic density evaluated for a given and fixed geometry $R$.} 
Assuming that they match for all $\lambda$ values leads to the following ground-state adiabatic density-functional approximation, according to the AC formula of Eq.~(\ref{eq:mol_AC_formula_Hxc_ener}),
\be\label{eq:Hxc_mol_DFA}
\mathcal{E}_{\rm
Hxc}[\Gamma,n]\approx \int dR\;\Gamma(R)\,E_{\rm Hxc}[n_R]. 
\ee
In practice, any (local or semi-local, for example) density-functional
approximation to $E_{\rm Hxc}[n]$ could be employed, thus turning the
theory into a practical computational method for beyond-BO computations. At this level of (adiabatic) approximation, the Hxc density-functional nuclear-electron and nuclear-nuclear potentials read (see Eqs.~(\ref{eq:Hxc_Ne_pot}) and (\ref{eq:Hxc_NN_pot}))
\be\label{eq:Ne_Hxc_pot_adia_approx}
V^{\rm
Hxc}_{\mathrm{ne}}[\Gamma,n](R,{\bf r})\approx\dfrac{\delta E_{\rm Hxc}[n_R]}{\delta n({\bf r})}
\ee
and
\be
\begin{split}
V^{\rm
Hxc}_{\mathrm{nn}}[\Gamma,n](R)&\approx E_{\rm Hxc}[n_R]
\\
&\quad-\int d{\bf r}\dfrac{\delta E_{\rm Hxc}[n_R]}{\delta n({\bf r})}n_R({\bf r}),
\end{split}
\ee
respectively. Interestingly, we recognize in the above expression the Levy--Zahariev (LZ) density-functional shift in
potential~\cite{levy2014ground},
\be\label{eq:LZ_shift_in_pot}
C_{\rm Hxc}[n]=\dfrac{\displaystyle E_{\rm Hxc}[n]-\int d{\bf r}\dfrac{\delta E_{\rm
Hxc}[n]}{\delta n({\bf r})}n({\bf r})}{\displaystyle\int d{\bf r}\;n({\bf r})},
\ee 
which enables to rewrite the exact ground-state energy of an electronic system as the sum of (shifted) occupied KS orbital energies. In the present context, the LZ shift ensures that, in the molecular KS Eq.~(\ref{eq:KS_mol_eq}), the total nuclear potential energy contains the electronic Hxc energy instead of its density-functional derivative, as expected. According to Eq.~(\ref{eq:LZ_shift_in_pot}), the adiabatic density-functional approximation to the nuclear-nuclear Hxc potential simply reads     
\be\label{eq:nn_Hxc_pot_adia_approx}
V^{\rm
Hxc}_{\mathrm{nn}}[\Gamma,n](R)&\approx \mathcal{N}_{\mathrm{e}}\;C_{\rm Hxc}[n_R].
\ee
\rev{To conclude, an adiabatic density-functional approximation to both Hxc nuclear-electron and nuclear-nuclear potentials has been derived (see Eqs.~(\ref{eq:Ne_Hxc_pot_adia_approx}) and (\ref{eq:nn_Hxc_pot_adia_approx})). Despite its simplicity, it is expected to provide non-trivial results when combined with the self-consistent BH-based formulation of molecular KS-DFT derived in Sec.~\ref{sec:mol-KS_DFT_adia_rep}. Indeed, while the zeroth iteration will consist in a conventional ground-state BO KS-DFT calculation, the excited KS states should become populated thanks to their non-adiabatic couplings with the ground KS state, thus updating the electronic density that is inserted into the Hxc potential (see Eqs.~(\ref{eq:eff_elec_dens_from_nuclear_and_elec_KS_wfs}) and (\ref{eq:Ne_Hxc_pot_adia_approx})). The practical implementation and calibration of the method is left for future work. }


\section{Conclusions and outlook}\label{sec:conclusions}

\rev{An alternative, general, and in-principle exact molecular KS-DFT has been derived. Its specificity is to map the fully quantum-mechanical molecular problem onto a fictitious (so-called KS) molecule where electrons interact with the nuclei but not among themselves anymore, thus leading to a drastically simplified KS molecular equation (Eq.~(\ref{eq:KS_mol_eq})) that must be solved self-consistently. When the KS molecular wavefunction is BH expanded (note that other mathematical forms like the exact factorization could also be employed in this context), the electronically non-interacting molecular problem turns into a ``KS beyond BO" electronic problem coupled to a nuclear one, where no additional step (like a response TD-DFT calculation) is needed to evaluate the non-adiabatic couplings between the different states}. (Quasi)diabatic representations could also be invoked in this context, upon minimizing the geometry-dependence of the ``KS beyond BO" electronic wavefunctions (Slater determinants or configuration-state functions), thus allowing for a simplified treatment of the action of the nuclear kinetic-energy operator as regards the electronic basis set, but at the expense of considering non-diagonal electronic matrix elements. This, obviously, paves an avenue for future work of much interest for molecular quantum dynamics.\\
Like in electronic KS-DFT, the Hxc energy, which describes the electron-electron repulsion within the molecule as a functional of both nuclear and (geometry-dependent) electronic densities, is the key quantity to approximate in molecular KS-DFT. An exact AC formula has been derived and, on that basis, we have proposed a simple adiabatic density-functional approximation that is compatible with any standard (local or semi-local) electronic DFT functional (see Eq.~(\ref{eq:Hxc_mol_DFA})). Such a strategy still needs to be assessed, which is left for future work.\\ 
Finally, in order to turn molecular KS-DFT into a practical computational method for non-adiabatic molecular quantum dynamics simulations, it should obviously be extended to the time-dependent regime, by analogy with time-dependent multi-component DFT~\cite{Butriy07}. Work is currently in progress in this direction.\\ 

\section*{Acknowledgements}
E.F. thanks ANR (CoLab project, grant no.: ANR-19-CE07-0024-02) for funding and the members of the PhotoMecha project (ANR-20-CE29-0021) for stimulating discussions. 



\appendix

\section{\revben{Electronic, nuclear, and molecular Hilbert spaces in the context of the nonadiabatic} Born--Huang expansion of the molecular wavefunction}\label{sec:mol_VP_BH_exp}


\revben{The objective of the present Appendix is twofold. First -- while its content is common knowledge for the readers belonging to the `nonadiabatic molecular dynamics' community --, we believe that a pedagogical recap may be useful for the readers belonging to the `ground-state electronic structure' community. Second, we focus here on clarifying the formal link between the restricted actions of electronic and nuclear operators within their own Hilbert spaces and their full action within the molecular Hilbert space. This aspect is often disregarded in nonadiabatic molecular dynamics because we end up in practice with well-defined closed-coupled equations for the nuclear wavefunction components after integration of the electronic wavefunctions over the electronic degrees of freedom. However, there are a few implicit shortcuts along the usual way of deriving them that may represent a source of practical confusion in the present context of molecular DFT, where the geometry-dependent one-electron density remains an explicit ingredient, together with the nuclear density.}

In what follows, we shall try to stay as general as possible as regards Dirac's ``bra and ket'' notations. The ``ket'' $\ket{\Psi}$ belonging to the molecular Hilbert space (for nuclei and electrons) corresponds to a wavefunction within the $(R,r)$-position representation that is expressed as
\be
\Psi(R,r)=\langle R,r\vert \Psi\rangle \quad.
\ee


\revben{The molecular Hamiltonian, $\hat{H}^{\rm
mol}$, and its various terms have been defined in the main text.} Except for $\hat{V}_{\rm
ne}$ (which explicitly creates a direct interaction between both electronic and nuclear Hilbert spaces), the other four operators can -- in principle -- be identified to their specific restrictions to their particular electronic or nuclear Hilbert spaces. However, we shall see in what follows that this deserves some precautions, especially for $\hat{T}_{\rm
n}$ \revben{when the electronic basis set shows some parametric dependence on the nuclear coordinates.}

\revben{Now,} according to the spirit of the adiabatic BO partition, we shall consider $R$ \revben{(the ``molecular geometry'', \ie, the set of translational-invariant nuclear coordinates; see main text)} as a given external parameter in the definition of the electronic wavefunctions and electronic Hamiltonian at each $R$,
\be
\hat{H}^{\rm
el}[R]=\hat{V}_{\rm
nn}[R]+\hat{F}_{\rm
e}+\hat{V}_{\rm
ne}[R], \forall
R \quad.
\ee
Such an operator is assumed to be restricted to acting within the electronic Hilbert space for a given value of $R$. Note that $\hat{V}_{\rm
nn}[R]$ acts as a mere energy shift as regards the electronic energy and is often removed temporarily. The corresponding ``Hamiltonian of quantum chemistry'' is thus
\be
\hat{H}^{\rm
qc}[R]=\hat{F}_{\rm
e}+\hat{V}_{\rm
ext}[R], \forall
R \quad,
\ee
where
\be
\hat{V}_{\rm
ext}[R]=\hat{V}_{\rm
ne}[R], \forall
R \quad,
\ee
is called the ``external potential'' in the language of DFT.
The electronic-only operator contribution restricted to the electronic Hilbert space,
\be
\hat{F}_{\rm
e} \equiv \hat{T}_{\rm
e}+\hat{W}_{\rm
ee} \quad,
\ee
describes the ``external-free'' many-body electronic problem and does not explicitly depend on $R$ as a parameter. In the language of DFT, it is the universal Hohenberg–Kohn (HK) term~\cite{Hohenberg1964}.

More rigorously, we should make the following identification \revben{within the molecular} Hilbert space,
\be
\hat{T}_{\rm
e}+\hat{W}_{\rm
ee} = \hat{\mathds{1}}_{\rm
n} \otimes\hat{F}_{\rm
e} \quad,
\ee
\revben{where $\hat{\mathds{1}}_{\rm
n}$ is the nuclear identity operator (acting within the nuclear Hilbert space). With this in mind, $\hat{T}_{\rm
e}+\hat{W}_{\rm
ee}$ is the trivial extension of $\hat{F}_{\rm
e}$ (acting within the electronic Hilbert space) to the molecular Hilbert space.}

\revben{The nuclear identity operator} can be expressed in various ways. As already alluded to, the $R$-position representation is to be preferred and will be somewhat unavoidable when defining $R$-parametrized electronic wavefunctions later on. Hence, 
\be
\hat{\mathds{1}}_{\rm
n}=\int dR\ket{R}\bra{R} \quad,
\ee
and
\be
\braket{R}{R'}=\delta(R-R') \quad,
\ee
where $\delta(\cdot)$ is Dirac's delta distribution.

We shall now introduce the notation $\hat{\mathds{1}}_{\rm
e}[R]$, which deserves some clarification. Let us consider that we assume $R$ as a fixed, external and continuously varying, parameter for defining the electronic problem (same spirit as in the BO approximation, but with no approximation yet). Then, we can express the resolution of the electronic identity operator (closure relationship) as
\be
\hat{\mathds{1}}_{\rm
e}[R]=\sum_\nu\ket{\Phi_\nu;R}\bra{\Phi_\nu;R}, \forall
R \quad,
\ee
such that we assume knowledge of $\{\ket{\Phi_\nu;R}\}_{\nu\in\mathbb{N}}$ for all $R$ as being a discrete, complete, orthonormal many-body electronic basis set, 
\be
\braket{\Phi_\nu;R}{\Phi_\mu;R}=\delta_{\nu\mu}, \forall
R \quad,
\ee
where $\delta_{\nu\mu}$ is the Kronecker symbol. Such a basis set (to be further specified later on) is assumed to span the electronic Hilbert space at each given value of $R$ and thus depends on it parametrically. In this, the semicolon notation is used to indicate the status of $R$ as being an external parameter and not a ``mechanical variable'' as regards the electronic Hilbert space. In practice, when working with spin-orbitals, this set spans a finite variational manifold made of, for example, all accessible Slater determinants (or configuration-state functions for a given total spin) according to some truncation scheme as regards configuration interaction. 

In the $r$-position representation, each electronic state corresponds to a many-body wavefunction, denoted
\be
\langle r\vert \Phi_\nu;R\rangle=\Phi_\nu(r;R), \forall
R \quad,
\ee
where
\begin{subequations}
\begin{align}
\label{eq:orthonormal_electronic_basis}
\braket{\Phi_\nu;R}{\Phi_\mu;R} =
\int dr\;  \Phi^*_\nu(r;R) \Phi_\mu(r;R) = \delta_{\nu\mu}, \forall
R \quad.
\end{align}
\end{subequations}
Electronic spin variables are implicit here but, in practice, they can be absorbed into $r$ if required (this is essential when dealing with Slater determinants generated by spin-orbitals but can be released eventually when invoking configuration-state functions constructed so as to span the subspace of some given spin multiplicity).

Obviously, if the basis set is complete for any $R$, the overall explicit dependence of $\hat{\mathds{1}}_{\rm
e}[R]$ with respect to $R$ vanishes, and we may want to drop $[R]$ and simply use $\hat{\mathds{1}}_{\rm
e}$ because the identity operator of a Hilbert space is unique. However, the choice of a given $R$-parametrized basis set creates an implicit dependence due to the fact that each basis vector in the expansion varies with $R$; this is an incentive for keeping $[R]$ as a reminder for this. Further, any practical calculation will eventually resort to a truncated basis set. Then, $\hat{\mathds{1}}_{\rm
e}[R]$ turns to be, in fact, a \revben{finite} projector onto the many-body Hilbert subspace (variational manifold), expected to vary explicitly, yet smoothly, with $R$. Interestingly enough, the first-order variation of such a subspace projector with $R$ occurs to be a manifestation of the so-called Pulay forces in analytic gradient techniques.

Although we know that such considerations are \revben{common} knowledge in the field of nonadiabatic dynamics and may seem over-pedantic, our target here is broader and we thus prefer to prevent any possible trap due to formal ambiguity, \revben{especially in the present context where we want to use one-electron and nuclear densities together.} 

Let us now turn to the two operators that depend explicitly on $R$ in the electronic Hamiltonian, namely $\hat{V}_{\rm
nn}[R]$ and $\hat{V}_{\rm
ne}[R]$. First, rewinding to $\hat{H}^{\rm
mol}$, we have, without any loss of generality, 
\be
\hat{V}_{\rm
nn}
=V_{\rm
nn}(\hat{R})\otimes\hat{\mathds{1}}_{\rm
e} \quad,
\ee
where $\hat{\mathds{1}}_{\rm
e}$ is, for the moment, free of any $R$-dependence. In this, $V_{\rm
nn}(\hat{R})$ is restricted to the nuclear Hilbert space and is local within the nuclear $R$-position representation, 
\be
\bra{R}V_{\rm
nn}(\hat{R})=V_{\rm
nn}(R)\bra{R} \quad.
\ee
Reconnecting to our previous definition of $\hat{H}^{\rm
el}[R]$, we thus obtain
\be
\hat{V}_{\rm
nn}[R]=V_{\rm
nn}(R)\hat{\mathds{1}}_{\rm
e}[R], \forall
R \quad,
\ee
which is now to be viewed indeed as a mere scalar ``energy shift'' operator acting in the electronic Hilbert space for any given value of $R$ (interestingly enough, it should be stressed that such notation subtleties are not inconsequential: by this, we have achieved a full transposition of $V_{\rm
nn}(\hat{R})$,  as an operator originally acting in the nuclear Hilbert space, to $\hat{V}_{\rm
nn}[R]$, as an operator that is a term \revben{present in} $\hat{H}^{\rm
el}[R]$ \revben{and absent in} $\hat{H}^{\rm
qc}[R]$, now acting in the electronic Hilbert space for all $R$).

The case of $\hat{V}_{\rm
ne}$ is perhaps more subtle, since it truly connects both the nuclear and electronic Hilbert spaces and cannot be factorized. Yet, because it is local within the $(R,r)$-position representation, we have 
\be
\bra{R,r}\hat{V}_{\rm
ne}=V_{\rm
ne}(R,r)\bra{R,r} \quad,
\ee
where $V_{\rm
ne}(R,r)=\sum^{\mathcal{N}_\mathrm{e}}_{i=1}V_{\rm ne}(R,{\bf r}_i)$. \revben{Note that we use the same notation $V_{\rm ne}(R,\cdot)$ for both the one-body and many-body functions for simplicity, since there is no ambiguity here. Further, its restriction to the electronic Hilbert space can be} denoted $\hat{V}_{\rm
ne}[R]=\hat{V}_{\rm
ext}[R], \forall
R$, as already mentioned, which is a term within either $\hat{H}^{\rm
el}[R]$ or $\hat{H}^{\rm
qc}[R]$, acting indeed in the electronic Hilbert space for all $R$.

Let us now return to the electronic basis set. It allows us to provide the following decomposition of the electronic Hamiltonian,
\be
\hat{H}^{\rm
el}[R]=\sum_{\nu,\mu}\ket{\Phi_\nu;R}
\mathcal{H}_{\nu\mu}(R)\bra{\Phi_\mu;R}
, \forall
R \quad,
\ee
where
\be
\mathcal{H}_{\nu\mu}(R)=\langle\Phi_\nu;R\left\vert\hat{H}^{\rm
el}[R]\right\vert\Phi_\mu;R\rangle, \forall R \quad.
\ee
Let us stress that the $\ket{\Phi_\nu;R}$ electronic states are not necessarily the so-called adiabatic (BO) \revben{eigensolutions} to the $R$-parameterized electronic Schr\"{o}dinger equation, which are those that specifically diagonalize the matrix $\mathbfcal{H}(R)=[\mathcal{H}_{\nu\mu}(R)], \forall R$.

We should now examine in detail the action of the nuclear kinetic-energy operator, $\hat{T}_{\rm
n}$. The $R$-position representation, formally denoted $\{\bra{R}\}_{R\in\mathbb{R}}$, satisfies
\be
\bra{R}\hat{R}=R\bra{R} \quad,
\ee
and
\be
\bra{R}\hat{P}=-\mathrm{i}\dfrac{\partial}{\partial R}\bra{R} \quad,
\ee
where $\hat{P}$ is the \revben{self-adjoint} momentum operator. It is such that the restriction of the nuclear kinetic-energy operator to the nuclear Hilbert space acts as
\be
\bra{R}T_{\rm
n}(\hat{P})=\bra{R}\dfrac{\hat{P}^2}{2M}=-\dfrac{1}{2M}\dfrac{\partial^2}{\partial R^2}\bra{R} \quad.
\ee

From the above considerations, we may now introduce the working expression of the resolution of the identity of the molecular Hilbert space as
\be
\hat{\mathds{1}}^{\rm
mol}=\int dR\ket{R}\otimes\hat{\mathds{1}}_{\rm
e}[R]\otimes\bra{R} \quad,
\ee
\ie, 
\be
\hat{\mathds{1}}^{\rm
mol}=\int dR\sum_\nu\ket{R}\otimes\ket{\Phi_\nu;R}\bra{\Phi_\nu;R}\otimes\bra{R} \quad,
\ee
hence,
\be
\hat{\mathds{1}}^{\rm
mol}=\int dR\sum_\nu\ket{R,\Phi_\nu;R}\bra{R,\Phi_\nu;R} \quad,
\ee
where
\be
\ket{R,\Phi_\nu;R}=\ket{R}\otimes\ket{\Phi_\nu;R}, \forall R \quad.
\ee
\revben{Let us recall here that $\langle r\vert\Phi_\nu;R\rangle = \Phi_\nu(r;R)$ is a basic wavefunction belonging to the electronic Hilbert space in $r$-position representation at some $R$, while $\langle R,r\vert R',\Phi_\nu;R'\rangle = \delta(R-R')\Phi_\nu(r;R')$ is a basic wavefunction belonging to the molecular Hilbert space in $(R,r)$-position representation. We may also use a partial $R$-representation in the molecular Hilbert space: $\langle R\vert R',\Phi_\nu;R'\rangle = \delta(R-R')\vert\Phi_\nu;R'\rangle$. In any case, it must be understood that we are fully bound to using the $R$-representation for the nuclear Hilbert space as long as we still have to explicitly invoke $R$-parameterized electronic states. Such precautions are rarely stressed out during typical derivations in the context of nonadiabatic molecular dynamics but can be found in, for example, Ref. \cite{mal14:1}.}

Now, the status of $\hat{T}_{\rm
n}$ is quite special and must be treated with care. Physically speaking, it is an operator that should be restricted to its action within the nuclear Hilbert space. More specifically, it reads 
\be
\hat{T}_{\rm
n}
=T_{\rm
n}(\hat{P})\otimes\hat{\mathds{1}}_{\rm
e} \quad,
\ee
whereby $T_{\rm
n}(\hat{P})$ is restricted to the nuclear Hilbert space and is nonlocal within the $R$-position representation (see above).

However, mathematically speaking, it also has a \revben{Riemannian differential} effect on the $R$-parameterized electronic basic \revben{vectors (``moving'' frame)}, which is the source of what are known as nonadiabatic couplings. \revben{Let us clarify this. From the above, we get
\be
\bra{R}\hat{T}_{\rm
n}
=-\dfrac{1}{2M}\dfrac{\partial^2}{\partial R^2}\left(\hat{\mathds{1}}_{\rm
e}[R]\otimes\bra{R}\right) \notag \\
=-\dfrac{1}{2M}\dfrac{\partial^2}{\partial R^2}\sum_\nu\ket{\Phi_\nu;R}\bra{\Phi_\nu;R}\otimes\bra{R} \notag \\
=-\dfrac{1}{2M}\dfrac{\partial^2}{\partial R^2}\sum_\nu\ket{\Phi_\nu;R}\bra{R,\Phi_\nu;R} \notag \\
=-\dfrac{1}{2M}\sum_\nu\dfrac{\partial^2}{\partial R^2}\left(\ket{\Phi_\nu;R}\bra{R,\Phi_\nu;R}\right), \forall R
\quad.
\ee
From the formula of the second derivative of a product, we obtain the partial $R$-representation of the full action of $\hat{T}_{\rm
n}$ in compact form as a sum of three terms, }
\be
\bra{R}\hat{T}_{\rm
n}
=-\dfrac{1}{2M}\sum_\nu
\ket{\Phi_\nu;R}\dfrac{\partial^2}{\partial R^2}\bra{R,\Phi_\nu;R}
\notag\\ 
+2\ket{\dfrac{\partial}{\partial R}\Phi_\nu;R}\dfrac{\partial}{\partial R}\bra{R,\Phi_\nu;R} 
\notag\\
+\ket{\dfrac{\partial^2}{\partial R^2}\Phi_\nu;R}\bra{R,\Phi_\nu;R}, \forall R \quad.
\ee
Such a notation is quite compact but perhaps not evident. It acts on a vector within the full Hilbert space and returns the partial $R$-representation of it. For better understanding, a \revben{total} representation in the full Hilbert space can be obtained upon also invoking the electronic \revben{basic vectors} as follows, 
\be
&\bra{R,\Phi_\nu;R}\hat{T}_{\rm
n}
=-\dfrac{1}{2M}\sum_\mu
\braket{\Phi_\nu;R}{\Phi_\mu;R}\dfrac{\partial^2}{\partial R^2}\bra{R,\Phi_\mu;R}
\notag\\ 
&+2\braket{\Phi_\nu;R}{\dfrac{\partial}{\partial R}\Phi_\mu;R}\dfrac{\partial}{\partial R}\bra{R,\Phi_\mu;R} 
\notag\\
&+\braket{\Phi_\nu;R}{\dfrac{\partial^2}{\partial R^2}\Phi_\mu;R}\bra{R,\Phi_\mu;R}, \forall R \quad,
\ee
where $\braket{\Phi_\nu;R}{\Phi_\mu;R}=\delta_{\nu\mu}$, and
\be
\braket{\Phi_\nu;R}{\dfrac{\partial}{\partial R}\Phi_\mu;R} = \mathcal{G}_{\nu\mu}^R(R) \quad,
\ee
\be
\braket{\Phi_\nu;R}{\dfrac{\partial^2}{\partial R^2}\Phi_\mu;R} = \mathcal{L}_{\nu\mu} (R) \quad,
\ee
are the gradient-component and Laplacian matrix-element contributions to the nonadiabatic coupling, respectively. \revben{Along the same line of formal precaution as before, we would like to stress here an important point: the $\mathbfcal{G}^R(R)$ and $\mathbfcal{L}(R)$ matrices are not the matrix representations of any operators acting within the electronic Hilbert space, but rather are the overlap matrices of electronic wavefunctions with their $R$-parametric derivatives (which is why we did not use the $\langle\cdot\left\vert\partial\right\vert\cdot\rangle$ notation but the $\langle\cdot\vert\partial\cdot\rangle$ one on purpose). Such a confusion is sometimes found in the literature.}

Hence, we obtain
\be
\begin{split}
&\bra{R,\Phi_\nu;R}\hat{T}_{\rm
n}
=-\dfrac{1}{2M}
\dfrac{\partial^2}{\partial R^2}\bra{R,\Phi_\nu;R}
\\ 
&-\dfrac{1}{2M}\sum_\mu \left(2\mathcal{G}_{\nu\mu}^R(R)\dfrac{\partial}{\partial R}
+\mathcal{L}_{\nu\mu}(R)\right) \bra{R,\Phi_\mu;R}, \forall R \quad.
\end{split}
\ee
The first term is the action of the nuclear kinetic-energy operator within the adiabatic BO approximation (no coupling). The rest (``coupled'' sum) is the action of the nonadiabatic coupling operator. Note that diagonal $\mathcal{L}_{\nu\nu}(R)$-terms are the so-called diagonal BO corrections (DBOC) to the potential energy, typically involved in what is sometimes called the adiabatic BH approximation (as opposed to the cruder BO approximation). 

In order to make things more concrete, we shall come back to the very definition of the molecular wavefunction, 
\begin{subequations}
\begin{align}
\braket{R,r}{\Psi}=\int dR'\sum_\nu\langle R,r \vert R',\Phi_\nu;R' \rangle \langle R',\Phi_\nu;R' \ket{\Psi} \quad,
\end{align}
\end{subequations}
\ie, 
\begin{subequations}
\begin{align}
\braket{R,r}{\Psi}=\int dR'\braket{R}{R'}\sum_\nu\langle r \vert \Phi_\nu;R' \rangle \langle R',\Phi_\nu;R' \ket{\Psi} \quad,
\end{align}
\end{subequations}
thus leading to the typical BH expansion,
\be
\Psi(R,r)=\sum_\nu\chi_\nu(R)\Phi_\nu(r;R)=\braket{R,r}{\Psi} \quad,
\ee
where
\be
\langle R, \Phi_\nu;R \ket{\Psi}=\chi_\nu(R) \quad.
\ee

From everything set above, we finally get \revben{an operator-matrix representation of the molecular Hamiltonian, $\mathbfcal{\hat{H}}^{\rm
mol}(\partial_R,R)$,  acting on the column-vector $\boldsymbol{\chi}(R)$ of nuclear wavefunction components,
\be
\begin{split}
&\sum_\mu\mathcal{\hat{H}}^{\rm
mol}_{\nu\mu}(\partial_R,R) \chi_\mu(R)
=\sum_\mu\Big(
-\dfrac{1}{2M}\delta_{\nu\mu}\partial^2_R
\\ 
&\quad- \dfrac{1}{M} \mathcal{G}_{\nu\mu}^R(R)\partial_R - \dfrac{1}{2M} \mathcal{L}_{\nu\mu}(R)
+\mathcal{H}_{\nu\mu}(R) \Big)\chi_\mu(R) \quad.
\end{split}
\ee
}

This is the general formulation that is used in \revben{BH-expanded \cite{bor54} nonadiabatic quantum dynamics (see, for example, Refs. \cite{dom04,dom11,las11:460})}. Let us note that we are now acting within the nuclear Hilbert space. As already mentioned, the adiabatic electronic representation is the one that makes $\mathbfcal{H}(R)$ diagonal, while (quasi)diabatic representations are supposed to make $\mathbfcal{G}^R(R)$ and $\mathbfcal{L}(R)$ vanish as much as possible for all $R$ (to make it short, the latter involve electronic wavefunctions, $\Phi_\nu(r;R)$, that vary parametrically as little as possible with $R$) \revben{\cite{bae06}
}.

\pagebreak




\newcommand{\Aa}[0]{Aa}

\end{document}